\documentclass[11pt,a4paper,oneside]{article}
\usepackage{graphicx,amssymb,amsmath,color,xcolor}
\usepackage[linkcolor={blue},citecolor={red},colorlinks=true]{hyperref}
\usepackage{enumerate}
\usepackage[margin=1.5 cm]{geometry}
\usepackage{slashed}
\usepackage{multirow}
\usepackage{cancel}
\usepackage[normalem]{ulem}
\usepackage{color}
\usepackage{physics}
\usepackage{feynmp-auto}
\usepackage{subfigure}

\def\beq{\begin{equation}}
\def\eeq{\end{equation}}
\def\bea{\begin{eqnarray}}
\def\eea{\end{eqnarray}}

\def\bit{\begin{itemize}}
\def\eit{\end{itemize}}

\def\l{\left}
\def\r{\right}

\def\ra{\rightarrow}

\def\baa{\begin{array}}
\def\eaa{\end{array}}

\def\simgt{\mathrel{\lower2.5pt\vbox{\lineskip=0pt\baselineskip=0pt
           \hbox{$>$}\hbox{$\sim$}}}}
\def\simlt{\mathrel{\lower2.5pt\vbox{\lineskip=0pt\baselineskip=0pt
           \hbox{$<$}\hbox{$\sim$}}}}

\def\bfc{\begin{figure}\begin{center}}
\def\efc{\end{center}\end{figure}}
\def\nn{\nonumber\\}

\begin{document}
 \begin{flushright}
DESY-25-082
\end{flushright}
\vspace{.2cm}
\begin{flushright}
\footnotesize
\end{flushright}
\color{black}

\begin{center}

{\huge \bf CP-violation in production of heavy neutrinos from bubble collisions}
\\

\medskip
\bigskip\color{black}\vspace{0.5cm}
\begin{centering}
{
{\large Martina Cataldi}$^{a}$,
{ \large Kristjan Müürsepp}$^{b}$ and
{\large Miguel Vanvlasselaer}$^{c}$
}
\\[7mm]

{\it \small $^a$ II. Institute of Theoretical Physics, Universität Hamburg, Luruper Chaussee 149, 22761 \& Deutsches Elektronen-Synchrotron DESY, Notkestr.\,85, 22607 Hamburg, Germany}

{\it \small $^b$ INFN, Laboratori Nazionali di Frascati, C.P. 13, 100044 Frascati, Italy \& 
Laboratory of High-Energy and Computational Physics, NICPB, R\"avala pst 10,  10143 Tallinn, Estonia }\\

{\it \small $^c$ Theoretische Natuurkunde and IIHE/ELEM, Vrije Universiteit Brussel, \& The  International Solvay Institutes, Pleinlaan 2, B-1050 Brussels, Belgium}

\end{centering}

\bigskip

\end{center}
\centerline{\bf Abstract}
\begin{quote}

First order phase transitions (FOPT) in the early Universe can be powerful emitters of both relativistic and heavy particles, upon the collision of ultra-relativistic bubble shells. If the particles coupling to the bubble wall have CP-violating interactions, the same collision process can also create a local lepton or baryon charge. This CP-violation can originate from different channels, which have only been partially addressed in the literature. We present a systematic analysis of the different channels inducing CP-violation during bubble collisions: 1) the decay of heavy particles 2) the production of heavy particles and 3) the production of light and relativistic Standard Model (SM) particles.  

As an illustration of the impact that such mechanisms can have on baryon number and dark matter (DM) abundance, we then introduce a simple model of cogenesis, separating a positive and a negative lepton number in the SM and a dark sector. The lepton number asymmetry in the SM can be used to explain the baryon asymmetry of the Universe (BAU), while the opposite asymmetry in the dark sector is responsible for determining the abundance of DM. 
Moreover, the masses of light neutrinos can be understood via the inverse seesaw mechanism, with the lepton-violating Majorana mass originating from the FOPT.

A typical signal produced by a FOPT is the irreducible gravitational wave (GW) background. We find that a substantial portion of the parameter space can be probed at future observatories like the Einstein Telescope (ET).

\end{quote}

\clearpage
\noindent\makebox[\linewidth]{\rule{\textwidth}{1pt}} 
\setcounter{tocdepth}{2}
\tableofcontents
\noindent\makebox[\linewidth]{\rule{\textwidth}{1pt}} 	

\newpage

\newpage

\section{Introduction}

In the era of multimessenger cosmology, phase transitions (PTs) occurring in the early Universe have enjoyed an ever-increasing focus from the particle physics community. PTs are associated with a change in the order parameter, which in particle physics is usually given by the vacuum expectation value (vev) of a scalar field, responsible for spontaneously breaking a symmetry of the theory. A particularly interesting case is a first order PT (FOPT) whereby the vev changes discontinuously across a barrier between the symmetric and broken phases. FOPTs proceed via the nucleation of spherically symmetric true vacuum bubbles in the false vacuum background. The nucleated bubbles then expand outwards until they collide with other bubbles, and the true vacuum bubbles fill the whole Universe. The physics of the FOPT is rather subtle. It entails a rich variety of phenomenological applications such as baryogenesis describing the observed BAU~\cite{Kuzmin:1985mm, Shaposhnikov:1986jp,Nelson:1991ab,Carena:1996wj,Cline:2017jvp,Long:2017rdo, Bruggisser:2022rdm,Bruggisser:2018mus,Morrissey:2012db,Azatov:2021irb, Huang:2022vkf, Baldes:2021vyz, Chun:2023ezg}, providing production mechanisms for heavy DM~\cite{Falkowski:2012fb, Baldes:2020kam,Hong:2020est, Azatov:2021ifm,Baldes:2021aph, Asadi:2021pwo, Lu:2022paj,Baldes:2022oev, Azatov:2022tii, Baldes:2023fsp,Kierkla:2022odc, Giudice:2024tcp,Gehrman:2023qjn} and primordial black holes~\cite{10.1143/PTP.68.1979,Kawana:2021tde,Jung:2021mku,Gouttenoire:2023naa,Lewicki:2023ioy}.  Complementarily to collider experiments, FOPTs featuring GW signals also provide an additional avenue for testing models of new physics~\cite{Witten:1984rs,Hogan_GW_1986,Kosowsky:1992vn,Kosowsky:1992rz,Kamionkowski:1993fg, Espinosa:2010hh, Athron:2023xlk}. An additional attractive feature of FOPTs is that, though not occuring in the SM, they can rather easily be realized in a large variety of well-motivated beyond the SM (BSM) models like composite Higgs~\cite{Bruggisser:2022rdm,Bruggisser:2018mus,Pasechnik:2023hwv, Azatov:2020nbe,Frandsen:2023vhu, Reichert:2022naa,Fujikura:2023fbi}, extended Higgs sectors~\cite{Delaunay:2007wb, Kurup:2017dzf, VonHarling:2017yew, Azatov:2019png, Ghosh:2020ipy,Aoki:2021oez,Badziak:2022ltm, Blasi:2022woz,Banerjee:2024qiu}, axion models~\cite{DelleRose:2019pgi, VonHarling:2019gme}, dark Yang-Mills sectors~\cite{Halverson:2020xpg,Morgante:2022zvc} and $B-L$ breaking sectors~\cite{Jinno:2016knw, Addazi:2023ftv}.  

The interactions between the bubble wall and the surrounding thermal plasma which we will refer to as BP interactions in what follows have recently been studied in great detail. In the regime of relativistic bubble expansion, when the boost factor grows very large, $\gamma_w  \equiv 1/\sqrt{1-v_w^2}\gg 1$ where $v_w$ denotes the velocity of the wall, it was first shown in~\cite{Bodeker:2017cim} 
that the ultra-fast bubble wall could allow exotic $1 \to 2$ interactions, otherwise forbidden in vacuum. Subsequently,~\cite{Azatov:2020ufh} argued that particles much heavier than the scale of 
the transition could be produced in $1 \to 1$ and $1 \to 2$ processes due to the Lorentz violating bubble wall background, which will subsequently propagate in shells around  
the bubble wall~\cite{Baldes:2024wuz}. The maximal mass of particles that can be produced via such interactions scales like $M^{\rm max}_{\rm BP} \sim \sqrt{\gamma_wvT_{\rm nuc}}$, where $T_{\rm nuc}$ is the nucleation temperature to be defined later and $v$ the scale of the symmetry breaking.

The later occurring collision of bubbles have been shown to also be
a powerful source of heavy particles, through the so-called bubble collision mechanism, that we will refer to as BC or the BC mechanism from now on, whereby particles as heavy as $M^{\rm max}_{\rm BC} \sim \gamma_w v$ can be produced upon the collision of bubble walls\cite{Falkowski:2012fb, WATKINS1992446, Giudice:2024tcp}. This mechanism of production has been studied for different phenomenological purposes. Its impact on the DM abundance was investigated in~\cite{Falkowski:2012fb, Giudice:2024tcp}, mainly in the context of heavy DM.  
Authors \cite{Katz:2016adq} discussed a model where the heavy particles that are created by the collision of ultrarelativistic bubbles decay out-of-equilibrium and violate the CP symmetry, contributing to the baryon number of the Universe. A similar setup was studied in \cite{Cataldi:2024pgt} and applied to baryogenesis via leptogenesis. 
In those two studies, the CP-violation was assumed to lie only in the decay of the heavy particles produced during the BC.  However, it is known that  in the context of the BP interactions, CP-violation in the \emph{production} of heavy state can lead to the dominant contribution to the final baryon number~\cite{Azatov:2021irb}. This CP-violation in production was neglected in former studies on the BC mechanism. 
Hence, in this paper, we extend the previous studies by computing the asymmetry resulting directly from the production of heavy particles, and we analyse the phenomenological impact.

Our study contributes to three different aspects:
\begin{itemize}
    \item First, we analyse the CP-violation in the production of the heavy particles in the case of the BC mechanism.
    \item Secondly, we study the CP-violation in the $1\to 3$ processes where the heavy particle is not produced on-shell, but the light particles from the SM are produced via an off-shell heavy particle. We will see that this process also contains CP-violation, and we will quantify it.
    \item Thirdly, we show the equivalence between the QFT formalism used in the studies of the BP interactions (see for example \cite{Azatov:2021irb}) and the formalism used in the literature on the BC mechanism~\cite{Falkowski:2012fb, WATKINS1992446, Giudice:2024tcp}. 
\end{itemize}

The remainder of this paper is organised as follows. In section \ref{sec:prod}, we review the mechanism of particle production via the collision of bubble walls and lay down a series of relevant expressions for the rest of the paper. In section \ref{sec:CP_viol}, we present the computation of the CP-violation during the BC. In section \ref{sec:Model}, we introduce a minimal model, taking advantage of the BC mechanism to produce the observed BAU and the DM abundance of the Universe while also providing a possible explanation for the masses of the light neutrinos via the seesaw mechanism. We finally evaluate the prospects for testing the model by GW observatories. Eventually, in section \ref{sec:conclusion}, we conclude. 

\section{Production of relativistic and heavy particles via bubble to bubble collisions}
\label{sec:prod}

In this section, we first review the production mechanism of the BC and present most of the necessary material for the rest of the study. We will consider two qualitatively different cases: in the first case the collision of the scalar shells of the true vacuum bubbles results in the production of heavy particles, while in the second case light boosted particles are produced instead. 

\subsection{General set-up}

As a first toy model, let us  augment the SM Lagrangian with the following terms
\bea 
\label{eq:main_model}
\mathcal{L} \supset  Y \phi  P_R\bar N  \chi + \frac{1}{2} m_N N \bar N + \frac{1}{2}m_\chi \chi \bar \chi +  \sum_{\alpha} y_{\alpha}P_R N (\tilde H \bar L_{\alpha}) - V(\phi,T),
\eea 
where $N,\chi$ are heavy Dirac fermions. In particular, $N$, that we will occasionally also refer to as a heavy neutrino, is absent from the plasma during the bubble wall collision, due to Boltzmann suppression. The lepton number assignments are as follows $L(N) = L(L_\alpha)= L(\chi) = 1$. We consider the following hierarchy of scales:
\bea 
\label{eq:hierofscales}
m_N \gg m_\chi \gg T_{\rm reh} \sim v \gg v_{\rm EW} \,,
\eea
where $\expval{\phi} = v$ is the scale of the symmetry breaking associated with the potential $V(\phi,T)$ and $v_{\rm EW}$ is the electroweak (EW) symmetry breaking scale. The reheating temperature $T_{\rm reh}$ denotes the temperature of the plasma after the PT has completed. At this level, the lepton number is a symmetry of the Lagrangian before and after the PT, and the field $\phi$ \emph{is not} charged under the lepton number. Conversely, the $\phi$ vev, associated to the minimum value of the potential $V(\phi, T)$ which we assume to induce a FOPT, could break some new global or gauge symmetry independently of the lepton number. We remain agnostic about the exact form of the potential for simplicity and generality.

Before moving on, let us briefly comment on the terms not included in our toy Lagrangian in Eq.\eqref{eq:main_model}. In principle, interactions of the type $\phi \Bar{N} N, \phi \chi \Bar{\chi}$ are also allowed by symmetries, however they are only subleading corrections to the masses of the heavy $N$ and $\chi$  in the broken phase, and do not impact the CP-violation in production.  Notice that the chirality assignments in Eq.\eqref{eq:main_model} are not the only possibility. Thus, we also study the opposite possibility in Appendix \ref{app:chiral_2}.  Finally, another interaction allowed without imposing additional symmetries is of the form $\Bar{L} H P_R \chi$. We will study the impact of such an operator at the end of section \ref{sec:asymmetry}. In Appendix \ref{app:other_real} we discuss a concrete model with a new local U(1) symmetry where such terms are avoided by construction.

\subsubsection{First order phase transitions and particle production }

Now let us connect our chosen toy model with the physics of FOPTs and the associated particle production mechanism. FOPTs proceed via the nucleation of true vacuum bubbles surrounded by the false vacuum. At the interface between the two phases, the scalar field $\phi$ interpolates between an unbroken $\expval{\phi} = v = 0$ and a broken phase $\expval{\phi} = v   \neq 0$. The energy difference $\Delta V$ between the two vacua corresponds to the latent heat released by the PT.
For strong PTs where vacuum energy dominates over thermal radiation $\Delta V$ can be parameterized via
\bea
    \Delta V \equiv V_{\expval{\phi}=0}- V_{\expval{\phi}=v} = c_V v ^4 \, .
\eea
After nucleation, the bubbles start to accelerate and can either i) reach a terminal velocity, or ii) enter the so-called runaway regime, accelerating closer and closer to the speed of light until collision. In the first case, most of the latent heat released by the transition goes to the formation of plasma sound waves and the heating of the plasma, thus limiting the energy available for particle production processes.  In the second case, a significant (order 1) fraction of the latent heat goes to the shear stress in the scalar wall, which accumulates energy until collision.

\noindent
Whether case i) or ii) is realized depends heavily on the model under consideration and depends on the forces acting on the bubble wall, which in turn control the bubble expansion velocity. These forces include, first, a driving force arising from the difference in vacuum energy between the false and true vacua:
\begin{equation}
\text{Driving force} = \Delta V \, ,
\end{equation}
as well as frictional forces due to interactions with plasma particles colliding with the bubble wall. In general, computing these effects is quite complex. However, in the highly relativistic regime $\gamma_w \gg 1$, the calculations simplify significantly. At tree level (leading order, LO), the pressure exerted by the plasma on the bubble wall~\cite{Bodeker:2009qy} takes the form:
\begin{equation}
\Delta \mathcal{P}_{\text{LO}} \to \sum_i g_i c_i \frac{\Delta m_i^2}{24} T_{\text{nuc}}^2 \, , 
\end{equation}
where $T_{\text{nuc}}$ is the nucleation temperature (i.e., the temperature at which the phase transition occurs), $\Delta m_i^2$ is the change in the squared mass of particle $i$ during the transition, and $c_i = 1$ for bosons ($1/2$ for fermions).
\noindent
When there is mixing between light and heavy particles, with coupling $Y$, a second leading-order (LO) friction term arises, originating from the mixing itself~\cite{Azatov:2020ufh}. For the models relevant to our analysis, this friction contribution takes the form:
\begin{equation}
\Delta \mathcal{P}^{\text{mixing}}_{\text{LO}} \to \frac{T^2 Y^2}{48 } v^2\Theta(\gamma_w T_{\text{nuc}} - M^2 L_w)  \, . 
\end{equation}
So, for pure scalar and fermionic theories with mild supercooling or gauged theories, since the two sources of pressure above saturate to a maximal value at large $\gamma_w$, one can conclude that if~\cite{Bodeker:2009qy}
\bea 
\Delta V > \Delta \mathcal{P}_{\text{LO}} + \mathcal{P}^{\text{mixing}}_{\text{LO}} \qquad \text{(Bodeker-Moore criterion)}\, ,
\eea 
i.e. the so-called \emph{Bodeker-Moore} criterion is satisfied,
the wall can runaway, and display large amounts of energy stored in the shear stress of the wall.

In the case of \emph{gauged theories},  due to the emission of soft gauge bosons, the so-called transition radiation effect, the friction on the wall scales linearly with the boost factor $\gamma_w \equiv 1/\sqrt{1-v_w^2}$~\cite{Bodeker:2017cim, Azatov:2020ufh, Gouttenoire:2021kjv, Azatov:2023xem, Ai:2025bjw}
\bea 
\mathcal{P}_{\text{transition radiation}} \sim \frac{\gamma g^3 v T^3}{16\pi^2}, \qquad \qquad \Rightarrow \qquad \qquad  \gamma_{\rm term} \sim \frac{16 \pi^2}{g^3} \bigg(\frac{v}{T_{\rm nuc}}\bigg)^3 \, . 
\eea 
If this $\gamma_{\rm term}$ is bigger than the boost factor at collision of bubbles $\gamma^{\rm coll
}_w$ (to be specified later, in Eq.\eqref{eq:gamma_coll}), then one can consider the bubble being \emph{effectively} in a runaway regime, it accelerating until collision. Requiring \emph{effective} runaway puts an upper bound on the gauge coupling $g$. For the numerical values that will be considered in this paper, we find, using Eq.\eqref{eq:gamma_coll} below and definitions around it, an upper bound $g_{\rm max} \sim (4\pi)^{2/3}\big(v^5/T_{\rm nuc}^4M_{\rm pl}\big)^{1/3} $, which lies in the range $g_{\rm max} \in [0.001, 1]$ for the values of $v/T_{\rm nuc} \in \left[1,10 \right]$ that are used in our numerical analysis. This corresponds to a regime of moderate supercooling.
\begin{figure}[ht]
    \centering
    \includegraphics[scale=0.33]{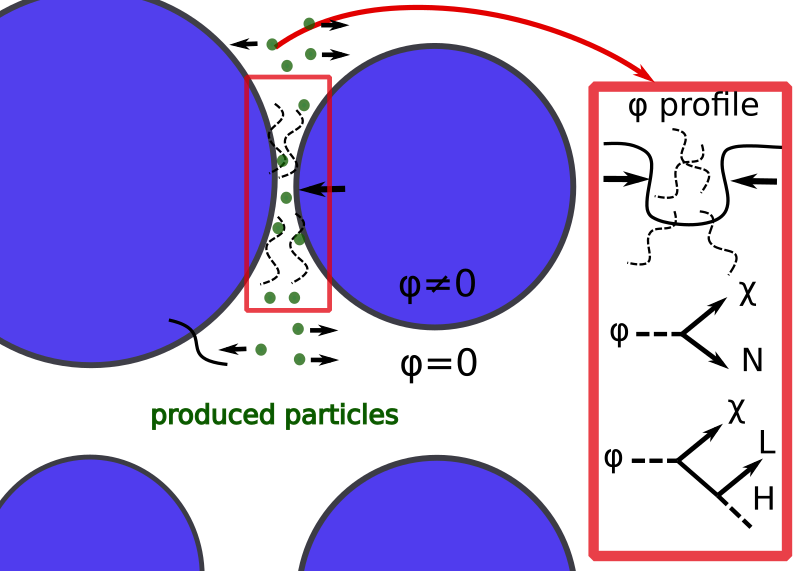}
    \includegraphics[scale=0.3]{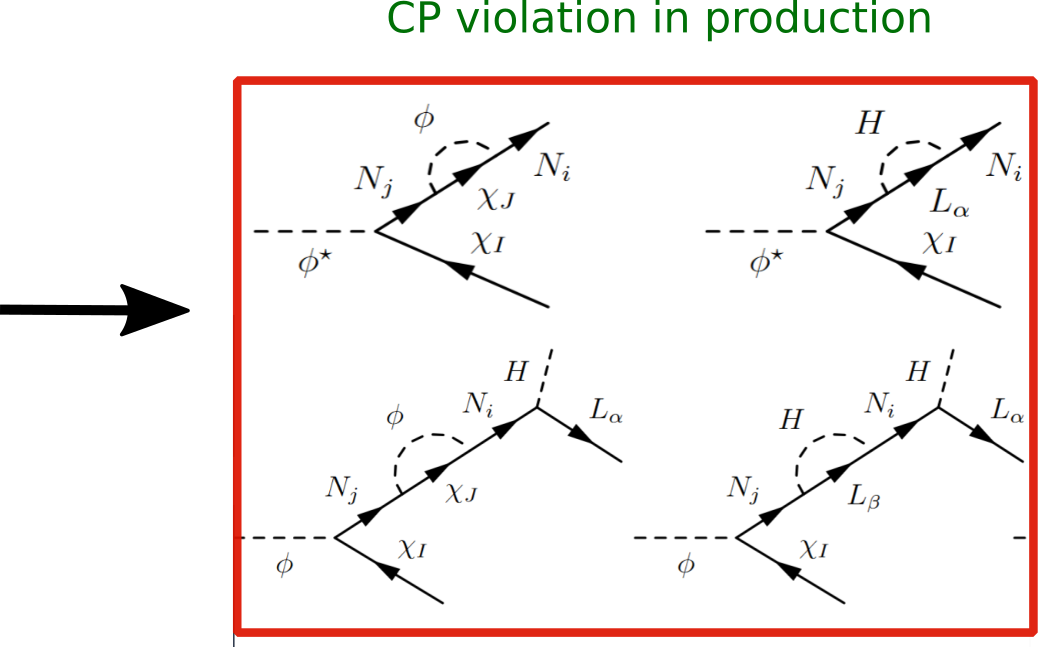}
    \caption{\textbf{Left Panel}: Cartoon of the particle production from the BC mechanism. The bubbles in blue collide, producing boosted particles via the channels $\phi^\star \to \chi N$ and $\phi^\star \to \chi HL$, with an off-shell $N$. \textbf{Right Panel}: Diagrams inducing the CP-violation via one loop insertion, the analytic form of those diagrams will be computed in this paper. }
    \label{fig:BtoBcollision}
\end{figure}

Finally, we briefly summarize the different particle production mechanisms that can be sourced by the PT dynamics, discussed above. First of all, a single bubble wall expanding in vacuum cannot produce particles. This can be intuitively understood by going to the wall frame where the scalar field profile is at rest, and no particle creation can thus occur. However, the interaction of a bubble with thermalized particles in the plasma violates the Lorentz symmetry and thus facilitates the production of both light and boosted particles~\cite{Bodeker:2017cim, Gouttenoire:2021kjv, Azatov:2023xem} as well as heavy particles~\cite{Azatov:2020ufh, Azatov:2021ifm, Azatov:2021irb}. Moreover, the Lorentz invariance violation during the collision of ultrarelativistic bubble walls provides another source for particle production independently of the presence of thermal plasma. In the remainder of this section, we will study the two CP-violating processes during the BC, that are encoded in the Lagrangian in Eq.\eqref{eq:main_model}: the production of a heavy fermion $N$, via $\phi^\star \to \chi N$ and the production of two light SM particles via $\phi^\star \to \chi HL$.   On Fig.\ref{fig:BtoBcollision}, we present an illustration of the process discussed in this paper, along with the intrinsic CP-violation.

\subsection{The production of a pair of heavy particles}

We start our investigation with the study of the process $\phi^{\star} \to \chi N$. 
The probability of emitting heavy particles is given by~\cite{Falkowski:2012fb, WATKINS1992446}~\footnote{The factor of $2$ in front of the integral comes from the fact that the probability of particle production from the dynamics of the scalar is given by the imaginary
part of its effective action, $\mathcal{P} = 2 \text{Im}[\Gamma]$. See \cite{Giudice:2024tcp} for more details. }

\begin{equation}
    P_{\phi \to  N \chi}= 2 \int \frac{dp_z d\omega}{(2\pi)^2} |\tilde \phi(p_z, \omega)|^2\text{Im}[\Sigma_{\phi \to  N \chi}(\omega^2 - p_z^2)] \, ,
\end{equation}
where $\tilde \phi(p_z, \omega)$ is the Fourier transform of the scalar profile and $\text{Im}[\Sigma_{\phi \to  N \chi}(\omega^2 - p_z^2)]$ denotes the imaginary part of the two point function $\expval{\tilde \phi (\omega^2 - p_z^2) \tilde \phi(\omega'^2 - p_z'^2)}$ given by the decay rate of $\phi$ via the optical theorem. It is given by
\begin{align}
\label{eq:rate_dec}
\text{Im}[\Sigma_{\phi \to  N \chi}(\omega^2 - p_z^2)] &= \frac{1}{2} \int \frac{d^3q d^3k}{(2\pi)^6 2E_k 2E_q }  \big|\mathcal{M}^2_{\phi \to \chi N}\big| (2\pi)^4 \delta^{(4)} (p-k-q)
\notag
\\ 
&= 
\frac{1}{16\pi}\frac{(\omega^2-p_z^2 - (m_{\chi}+ m_N)^2)^{3/2}}{\sqrt{\omega^2 - p_z^2}}  \abs{Y}^2 \theta(\omega^2-p_z^2 - (m_{\chi}+ m_N)^2) 
\, . 
\end{align} 
The probability of production is computed via two methods in Appendix \ref{app:production}, where we show that the method used 
in Ref.~\cite{Azatov:2021irb} to compute the production rate from the BP interactions (AVY method) is equivalent to the computation in Ref.~\cite{WATKINS1992446} (WW), used for the rate in the context of the BC mechanism.  Let us note that the result in the main text has a further factor of $1/2$ compared to the result in the Appendix because of the additional projector on the right-handed degree of freedom present in the Lagrangian Eq.\eqref{eq:main_model}.

Consequently, in the context of planar walls, the number of heavy particles per surface area produced during the BC is given by  
\begin{equation}
\label{eq:prod_heavy_N}
\frac{N}{A}\bigg|_{ N } =\frac{N}{A}\bigg|_{ \chi } =
\int \frac{ dp_z d\omega}{(2\pi)^2 8\pi } \frac{(\omega^2-p_z^2 - (m_{\chi}+ m_N)^2)^{3/2}}{\sqrt{\omega^2 - p_z^2}}  \abs{Y}^2 \theta(\omega^2-p_z^2 - (m_{\chi}+ m_N)^2)  |\tilde\phi( p_z, \omega)|^2 \,. 
\end{equation}

At the locus of the bubble wall collision, the Fourier transform of the double wall becomes non-trivial. We parameterize it by defining a function  $f(p^2)$
  \bea 
  f(p^2) \equiv |\phi(\omega^2-  p_z^2)|^2 \, ,
  \eea 
  which has been computed in the literature~\cite{Falkowski:2012fb, Giudice:2024tcp} and is reported in Appendix \ref{app:f_funct}. The number of produced particles per unit surface becomes
\bea 
\label{eq:emission_N}
\frac{N}{A}\bigg|_{ N }=
\frac{\abs{Y}^2}{2\pi^2}\frac{1}{16\pi}\int^{p_{\rm max}^2}_{(m_{\chi}+ m_N)^2} dp^2 \frac{(p^2 - (m_{\chi}+ m_N)^2)^{3/2}}{\sqrt{p^2}}      f(p^2)  \, . 
\eea 

The upper cut-off $p_{\rm max}^2$ is introduced by hand and is given by the thickness of the wall in the plasma frame $ \sim (2 \gamma_w/L_w)^2$, where $L_w \sim 1/v$.  After the collision of the bubbles has finished, the particles produced by the collision diffuse over the radius of the bubble, homogenizing their distribution\footnote{The heavy particles could of course, decay before diffusing. We will consider this effect later on in section \ref{sec:Model}.}.
At the end of this process, the yield of heavy particles is given by 
\begin{align} 
 Y_N \equiv\frac{n_N}{s} = \frac{n_\chi}{s} &= \frac{1}{s(T_{\rm reh})}\overbrace{\frac{N}{A}\bigg|_N}^{\text{production per surface}}\times \overbrace{\frac{3}{2 R_{\rm coll}}}^{\text{diffusion}}
 \notag \\
 &\approx \frac{1}{s(T_{\rm reh})}\frac{3 \beta H }{4 (8 \pi)^{4/3}v_w }\times \frac{\abs{Y}^2}{2\pi^2}\int^{p_{\rm max}^2}_{(m_{\chi}+ m_N)^2} dp^2 \frac{(p^2 - (m_{\chi}+ m_N)^2)^{3/2}}{\sqrt{p^2}}     f(p^2),
 \label{eq:2heavyyieldfull}
 \end{align} 
 where $R_{\rm coll}$ is the average radius of the bubbles at collision and the entropy density $s$ is given by $
s(T) \equiv g_\star \frac{2\pi^2}{45}T^3
$ with  $g_\star$ denoting the number of relativistic degrees of freedom.  The radius of the bubbles at collision is controlled by the typical dimensionless duration of the transition $\beta(T)$
\bea
 R_{\rm coll}\approx \frac{(8 \pi)^{1/3}v_w}{H(T_{\rm reh})\beta(T_{\rm reh})},~~~\beta(T)= T\frac{d}{dT}\l(\frac{S_3}{T}\r) \, , 
\eea
where $v_w$ denotes the speed of the expanding wall and the Hubble rate is given by
\bea 
H^2(T) = \frac{\rho_{\rm rad}}{3 M_{\rm Pl}^2} = \frac{1}{3 M_{\rm Pl}^2}\frac{\pi^2 g_\star}{30}T^4 \, ,
\eea 
where $M_{\rm Pl} \approx 2.4 \times 10^{18}$ GeV is the reduced Planck mass 
and $S_3$ is the $O(3)$-symmetric Euclidean action of the nucleating bubble, obtained by solving the bounce equation for the scalar profile, which determines the rate of the PT. The nucleation temperature $T_{\rm nuc}$ can be estimated from $S_3$ via
\begin{equation}
    \frac{S_3}{T_{\rm nuc}} \simeq 4 \log \left( \frac{T_{\rm nuc}}{H(T_{\rm nuc})} \right) + \frac{3}{2} \log \left( \frac{S_3}{2 \pi T_{\rm nuc}} \right) \, .
\end{equation}

Using energy conservation, both $T_{\rm reh}$ and $T_{\rm nuc}$ can be expressed as
\begin{equation}
 T_{\rm reh} = \bigg( \frac{30 (1+ \alpha)}{g_\star \pi^2 \alpha} c_V\bigg)^{1/4} v  \,, \hspace {8mm} T_{\rm nuc} = \bigg( \frac{30}{g_{\star i} \pi^2 \alpha} c_V\bigg)^{1/4} v \, .
\end{equation}
where $g_\star$, $g_{\star,i}$ are the number of relativistic degrees of freedom at $T=T_{\rm reh}$, $T_{\rm nuc}$ respectively with $ g_\star = g_{\star i}=106.75$, and $\alpha$ the strength of the PT, determined by the ratio of the energy released in the transition to the thermal energy of the plasma $ \alpha \equiv \Delta V/\rho_{\rm rad}$.  

After the PT completes, the Universe is filled by the true vacuum, while the rest of the latent heat is relased to the plasma, which heats up, increasing the entropy. To obtain the yield $Y_N \equiv n_N/s$  one can perform an order of magnitude estimate:
\bea 
\label{eq:2heavyparticles_estimate}
\int^{(2\gamma_w v)^2}_{(m_{\chi}+ m_N)^2} dp^2 \frac{(p^2 - (m_{\chi}+ m_N)^2)^{3/2}}{\sqrt{p^2}}     f(p^2) \approx N \times 16 v^2\int^{(2\gamma_w v)^2}_{(m_{\chi}+ m_N)^2} dp^2 \frac{(p^2 - (m_{\chi}+ m_N)^2)^{3/2}}{p^5}       \, , 
\eea 
where N denotes the approximation for the p-dependent logarithm in the f-function, $2 \log{(\sqrt{z^2-1}+z)} \approx N$, where $z \equiv 2\gamma_w v/p$, as explained in Appendix \ref{app:f_funct}. We will keep using this expression in the rest of the paper.

An important comment is 
now in order: before the results of~\cite{Mansour:2023fwj} it was believed that only \emph{elastic} collisions could generate an appreciable abundance of very heavy particles. In~\cite{Mansour:2023fwj}, however, it was shown numerically that both elastic and inelastic collision share the same dominant contribution term $f_{\rm PE}$ for the production, making the particle production almost independent of the type of the collision. As a conservative model-independent assumption, in the following we will only consider the contribution from perfectly elastic collisions, neglecting the oscillation around the peak that was also found in~\cite{Mansour:2023fwj}.

Combining Eqs. \eqref{eq:2heavyyieldfull} and \eqref{eq:2heavyparticles_estimate} the yield of heavy particles $N$ due to BC will scale like\footnote{Notice that this result differs by a factor $ 1/3.2$ with respect to the one presented in \cite{Cataldi:2024pgt}. A factor of $1/2$ is explained by the fact that we have accounted for a missed factor of $1/2$ in Eq.\eqref{eq:rate_dec}.}
\bea 
\label{eq:Y_CB_app}
 Y_N^{\rm BC} \simeq  0.012 N \abs{Y}^2 \frac{\beta }{ v_w} \bigg( \frac{\pi^2 \alpha}{30(1+\alpha) g_\star  c_V}\bigg)^{1/4}\frac{v}{M_{\rm Pl} } \log \left( \frac{2 \gamma_w v}{m_\chi + m_N} \right)\, ,  
\eea

We observe immediately that, except for the cases of very large $v$, the produced abundance $Y^{\rm BC}_{N}$ is much smaller than the abundance of light particles in the plasma, which is typically $Y_{\rm light} \sim 10^{-3}$.

\subsubsection{A possible saturation and implementation of the UV cut-off}
\label{sect:UV_cutoff}

For consistency, it is important to check that the energy of the particles produced via the BC mechanism does not exceed the overall energy released by the PT, otherwise the production would backreact on the wall dynamics and likely suppress the resulting number density.  Before any rescattering with the plasma, we can estimate the energy in the produced particles to be\footnote{Notice that we did not include the dilution factor due to the injection of entropy into this expression. This is because we are  computing the energy density \emph{injected in the daughter particles}, to study the possibility of a backreaction on the collision process. Consequently, we need to consider this energy density before any dilution. }

\begin{equation}
\frac{E}{A}\bigg|_{ N } \approx \frac{E}{A}\bigg|_{ \chi } \approx \frac{1}{2}
\int \frac{ dp_z d\omega}{(2\pi)^2 8\pi } \omega \frac{(\omega^2-p_z^2 - (m_{\chi}+ m_N)^2)^{3/2}}{\sqrt{\omega^2 - p_z^2}}  \abs{Y}^2 \theta(\omega^2-p_z^2 - (m_{\chi}+ m_N)^2)  |\tilde \phi(\omega^2- p_z^2)|^2 \,. 
\end{equation}
The overall factor of $1/2$ is added to take into account the fact that more or less half of the energy goes into each particle. 
After diffusion of the fast particles, the energy density becomes 
\bea 
\rho_i \approx \frac{3 \beta H }{4 (8 \pi)^{4/3}v_w} \times \frac{E}{A}\bigg|_{ i } \, . 
\eea
In our model, those expressions can be estimated to give

 \bea 
 \rho^{\rm BC}_N = \rho^{\rm BC}_\chi \approx \frac{1}{2}\frac{3 \beta H }{4 (8 \pi)^{4/3}v_w}\times \frac{\abs{Y}^2}{2\pi^2}\int^{p_{\rm max}^2}_{(m_{\chi}+ m_N)^2} dp^2 \bigg(p^2 - (m_{\chi}+ m_N)^2\bigg)^{3/2}      f(p^2) \, .
 \label{eq: rhoN saturation}
 \eea
 Using Eq.\eqref{eq: rhoN saturation} the energy density stored in the heavy particles at the production can be then estimated as

\bea
\rho_N^{\rm BC} = \rho_\chi^{\rm BC} \approx 0.0027 N \abs{Y}^2 \frac{\beta}{v_w} \frac{\sqrt{g_\star} T_{\rm reh}^2 v^2}{M_{\rm Pl}} p_{\rm max}
\eea

This energy density is a linear function of the wall boost factor, and thus the energy of produced particles may grow without limit unless a cut-off is introduced. The boost factor for runaway bubbles at collision is given by 
\bea 
\label{eq:gamma_coll}
\gamma(R) \propto \frac{2}{3}\frac{R}{R_{\rm nuc}} \sim \frac{2}{3}RT_{\rm nuc} \qquad \Rightarrow
\gamma^{\rm coll
}_w \sim \frac{2 \sqrt{10} M_{\rm Pl} T_{\rm nuc} (8\pi)^{1/3} v_w }{ \pi \sqrt{g_\star} \beta T_{\rm reh}^2} \approx 5.9\frac{ M_{\rm Pl} T_{\rm nuc} v_w}{ \sqrt{g_\star} \beta T_{\rm reh}^2} \, .
\eea 
Using Eq.\eqref{eq:gamma_coll}, the energy density in heavy particles becomes
\bea 
\rho^{\rm BC}_N \approx 0.016 N|Y|^2 v^3T_{\rm nuc}  \, .  
\eea 
Neglecting backreactions of the particle production on the wall motion would require 
 \bea 
 \label{eq:backreact_cond}
  \rho^{\rm BC}_N + \rho^{\rm BC}_\chi \ll \Delta V \, ,
 \eea 
The opposite regime $\rho^{\rm BC}_N + \rho^{\rm BC}_\chi \sim \Delta V$ denotes the regime of strong backreaction of the emission on the wall. \\ \\

As a conjecture, let us now explain how to implement a cut-off to ensure that the energy of the particles produced by the BC mechanism does not exceed the energy released by the PT, for all of parameter space. To make this requirement more concrete, we define an upper cut-off $\mu_M$ by replacing the upper bound in the integral of Eq.\eqref{eq: rhoN saturation} $p_{\rm max} \to \mu_M$\, 
\bea 
 \rho^{\rm BC}_N(\mu_M) = \rho^{\rm BC}_\chi(\mu_M) \approx \frac{1}{2}\frac{3 \beta H }{4 (8 \pi)^{4/3}v_w}\times \frac{\abs{Y}^2}{2\pi^2}\int^{\mu_M^2}_{(m_{\chi}+ m_N)^2} dp^2 \frac{p}{2} \frac{\Big(p^2 - (m_{\chi}+ m_N)^2\Big)^{3/2} }{\sqrt{p^2}}     f(p^2) \, .
 \label{eq: rhoN saturation bis} 
\eea 
The value of the upper cut-off is then set by 
\bea 
2\rho^{\rm MAX}_N(\mu_M) = \Delta V \, ,  
\eea 
which can be solved for $\mu_M$. The rationale behind this cut-off is that the production of particles, when it starts to backreact on the wall shape, is naively expected to make the wall thicker, which decreases the upper bound in the integral set by the thickness of the wall. 
Let us now estimate the effect of the backreaction on the number of emitted particles. When we implement this same cut-off for the density produced, one obtains 
\bea 
\label{eq:Final_expression}
 Y_N^{\rm BC} \simeq  0.012 N \abs{Y}^2 \frac{\beta }{ v_w} \bigg( \frac{\pi^2 \alpha}{30(1+\alpha) g_\star  c_V}\bigg)^{1/4}\frac{v}{M_{\rm Pl} } \log \left( \frac{\text{Min}[2 \gamma_w v, \mu_M]}{m_\chi + m_N} \right)\, .  
\eea

We conclude that the backreaction effects are only expected to lead to a logarithmic correction to the number of particles produced. This is only an order one correction if $\mu_M \gg m_N+ m_\chi$. We will use Eq.\eqref{eq:Final_expression} in the remainder of this paper.

\subsection{Emission of light boosted particles}

Due to the hierarchy outlined in Eq.\eqref{eq:hierofscales}, the $N$ produced during BCs will almost immediately decay either to $\phi \chi$ or to $HL$. Importantly, however, for such production, $N$  does not necessarily have to be \textit{on-shell}. In particular, if the scalar shells of the bubble walls are not boosted enough, i.e., if $\gamma_w v \ll m_N$, the production of $\chi H L$ or $\chi \phi$ can also occur via \textit{off-shell} $N$. In this section, we make use of that observation to specifically investigate the emission of light boosted particles via an \textit{off-shell} $N$. In this case, the dominant interaction is  $\phi^\star \to \chi HL$. The three-body decay rate, for $HL\chi$, is given by 
\begin{equation}
 \Gamma_{\phi^\star \to HL \chi}(p^2) = \frac{2}{(2 \pi)^3} \frac{ \abs{y}^2 \abs{Y}^2}{32 \sqrt{p^6} } \int_{s_{12}^{\text{min}}}^{s_{12}^{\text{max}}} \int_{s_{23}^{\text{min}}}^{s_{23}^{\text{max}}} \frac{m_N^2 (s_{23} - m_L^2 - m_{\chi}^2) }{(s_{12} - m_N^2)^2  + m_N^2 \Gamma_N^2} \text{d} s_{12} \, \text{d} s_{23}  \, ,
 \label{eq:1gto3}
\end{equation}
containing a sum over the outgoing spins and where $\Gamma_N$ denotes the total decay width of $N$. The integration limits for $s_{23}$ are given by 
\begin{align}
    s_{23}^{\text{max}} &= (E_L^{\star} + E_{\chi}^{\star})^2 - \left(\sqrt{(E_L^{\star})^2 - m_L^2} - \sqrt{(E_\chi^{\star})^2 - m_\chi^2}\right)^2, \\
    s_{23}^{\text{min}} &= (E_L^{\star} + E_{\chi}^{\star})^2 - \left(\sqrt{(E_L^{\star})^2 - m_L^2} + \sqrt{(E_\chi^{\star})^2 - m_\chi^2}\right)^2,
    \label{eq:s23lim}
\end{align}
with
\begin{equation}
E_{L}^{\star} = \frac{s_{12} - m_H^2 + m_L^2}{2\sqrt{s_{12}}}, \hspace{0.5cm}  E_{\chi}^{\star} = \frac{p^2 - s_{12} -m_{\chi}^2}{2 \sqrt{s_{12}}} \, . 
\end{equation} 
The integration limits over $s_{12}$ are 
\begin{equation}
s_{12}^{\text{min}} = (m_H + m_L)^2,   \hspace{0.5cm} s_{12}^{\text{max}} = (p-m_{\chi})^2.
\end{equation}
In our regime of interest $m_N > m_\chi \gg m_L, m_H$ the asymptotic behaviour of the integral at low and high energies can be described by simple analytic formulae: 
 \begin{align}
 \text{\textbf{Low energies}:} \qquad  & p \ll m_{N}: \qquad \qquad   \Gamma_{\phi^\star \to HL \chi} \simeq 2\frac{\abs{y}^2 \abs{Y}^2}{1536 \pi^3} \frac{p^3}{m_N^2} 
  \\
\text{\textbf{High energies}:} \qquad  & p \gg m_{N}: \qquad \qquad \Gamma_{\phi^\star \to HL \chi} \simeq  2\frac{\abs{y}^2 \abs{Y}^2p}{512\,\pi^2}\frac{m_N}{\Gamma_N}  \, . 
\end{align}
 We have explicitly checked the validity of these analytic estimates against the numerical evaluation of Eq.\eqref{eq:1gto3}.
Let us now study those two regimes separately. 

\subsubsection{Emission of light particles at low energies}

Assuming $m_L, m_H \ll p \ll m_N$, we now compute the number density and energy density of the emitted $HL$ pair (denoted by $N_{\rm SM}$ and $\rho_{\rm SM}$ respectively), via an expression analogous to       Eq.\eqref{eq:emission_N}
\begin{align}
\frac{N_{H,L,\chi}}{A}\bigg|^{}_{\phi^\star \to \chi HL} 
&\approx 2\frac{\abs{y}^2 \abs{Y}^2 }{1536 \pi^3 } \int^{p_{\rm max}^2}_{m_\chi^2} \frac{dp^2}{2 \pi^2} \frac{ p^4}{ m_{N}^2} f(p^2) \, , 
\\
\Rightarrow \qquad  & n_{H,L,\chi} \approx \frac{3 \beta H }{ (8 \pi)^{1/3}v_w }\times \frac{\abs{y}^2 \abs{Y}^2 }{1536 \pi^3 } \int^{p_{\rm max}^2}_{m_\chi^2} \frac{dp^2}{2 \pi^2  } \frac{ p^4}{ m_{N}^2} f(p^2) \, . 
\end{align} 
Following the same logic as in the previous subsection, one can estimate the yield of light boosted $LH$ pair and $\chi$ produced by the bubbles in the plasma to be 
\begin{align}
\label{eq:YSM_light_low}
Y^{\rm BC}_{H, L,\chi} &\approx \frac{1}{s(T_{\rm reh})}\frac{3 \beta H}{ (8 \pi)^{1/3}v_w }\times   \frac{p_{\rm max}^2}{ m_{N}^2 } \frac{N \times 16 \abs{y}^2 \abs{Y}^2 }{1536 \pi^3 (2\pi^2) } v^2  
\notag 
\\
&\sim    1.3 \times 10^{-5} N \abs{y}^2 \abs{Y}^2\frac{\beta }{ v_w} \bigg( \frac{\pi^2 \alpha}{30(1+\alpha) g_\star  c_V}\bigg)^{1/4} \frac{p_{\rm max}^2v}{ m_{N}^2 M_{\rm Pl} } \, .  
\end{align}
Similarly, the energy density is given by 
\bea 
\label{eq:rho_light}
\rho_{H,L,\chi} = \rho_{L}+ \rho_{H} + \rho_{\chi}\sim 4 \times 10^{-6} N \abs{y}^2 \abs{Y}^2  \frac{\beta }{ v_w} \bigg( \frac{ 30(1+\alpha) c_V}{\pi^2 \alpha}\bigg)^{1/2} \frac{1 }{M_{\rm Pl}} \frac{p_{\rm max}^3v^4}{ m_{N}^2 } \, , 
\eea 
where both equations are valid in the limit $p_{\rm max}^2 \ll m_N^2$. Thus, we see that the number density and the energy density from the light SM states is suppressed by $p_{\rm max}^2/m_N^2$ compared to the energy density released by the PT. Consequently, backreaction effects can be safely neglected.

\subsubsection{Emission of light particles at high energies}

On the other hand, when $p \gg m_N$, the number density of emitted particles for the process $\phi^\star \rightarrow \chi H L$ can be written as 
\bea 
\label{eq:factorisation}
\frac{N_{L, H, \chi}}{A}\bigg|^{}_{\phi^\star \to \chi HL} 
\approx  2\int \frac{dp^2}{(2\pi)^2  } \frac{\abs{y}^2 \abs{Y}^2 p^2}{512\,\pi^2}\frac{m_N}{\Gamma_N}  f(p^2) = \bigg(\frac{\Gamma_{N \rightarrow L H}}{\Gamma_N}\bigg) \times \underbrace{\frac{1}{ 2}\frac{N_N}{A}\bigg|^{}_{\phi^\star \to \chi N}}_{\text{Production rate averaged over final N spins}}
\eea 
with the total decay rate $\Gamma_N = 2\Gamma_{N \rightarrow L H}+2\Gamma_{N \rightarrow \chi \phi}$ and $\Gamma_{N \rightarrow L H}= |y|^2 m_N/(16 \pi)$. We can interpret the ratio 
\bea 
\bigg(\frac{\Gamma_{N \rightarrow L H}}{\Gamma_N}\bigg) \equiv \text{Br}[N \to HL] \,,
\eea 
as the branching ratio of the decay of $N$ to the light SM species. Notice the factor of $1/2$ in front of the production $\phi^\star \to N\chi$, which can be understood from the fact that $N_N/A\big|^{}_{\phi^\star \to \chi N}$ is the rate of production of $N$ \emph{summed} over final states. To correct for this, we average over the spin of the final state $N$-s.  

This result hints that for energies larger than the resonance, the production of light $HL$ is largely dominated by the creation of \emph{on-shell} fermion $N$, also in the case of very large energies.

\section{The asymmetry produced by the bubble collision}
\label{sec:CP_viol}

\subsection{Baryon asymmetry of the universe (BAU)}

One of the most interesting open questions in modern cosmology is the observed asymmetry between matter and antimatter, the so-called BAU.

BAU can be expressed as the net comoving number density of baryons. The numerical value, obtained by the latest measurements of the cosmic microwave background (CMB)~\cite{Planck:2015fie} and the primordial abundance of light elements~\cite{Fields:2019pfx}, is given by
\bea 
\label{eq:observed_BAU}
Y_{\Delta B} \equiv \frac{n_{B}-n_{\overline{B}}}{s} \bigg|_{0} \approx (8.69 \pm 0.22) \times 10^{-11}
\eea 
where $n_{B}$, $n_{\overline{B}}$ and $s$ denote the number densities of baryons, antibaryons and entropy at present.
\begin{figure}[h!]
\centering
\includegraphics[scale=0.45]{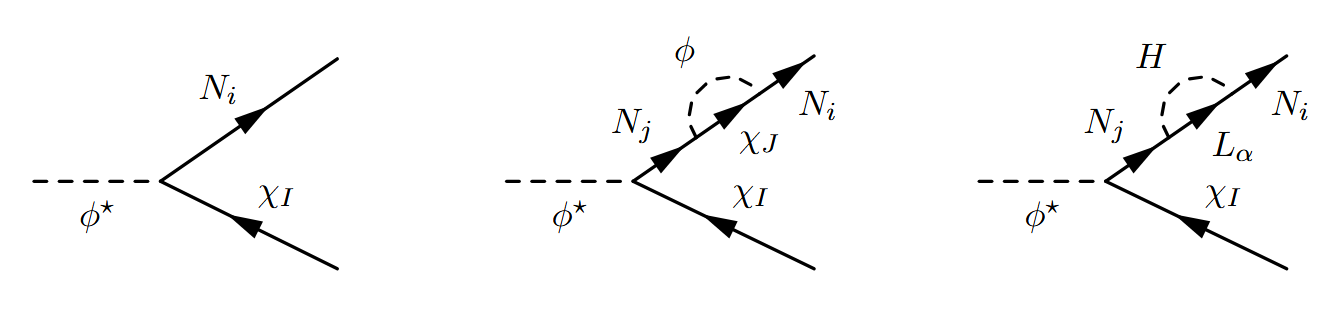}
\label{fig:CP_viold}
\caption{Feynman diagrams for the CP-violation in the production of the $N_i$ and $\chi_I$. }
\end{figure}
The observed value of BAU cannot be simply interpreted as an initial condition in the very early Universe, because inflation would dilute away any pre-existing asymmetry.
Instead, the value given by Eq.\eqref{eq:observed_BAU} needs to be generated during the evolution of the Universe via the so-called `baryogenesis' mechanism occurring after inflation.  For a successful baryogenesis mechanism, the three Sakharov requirements need to be satisfied~\cite{Sakharov:1967dj}:  i) violation of Standard Model (SM) baryon number, ii) violation of charge ($C$)
and charge-parity ($CP$) symmetries, and iii) departure from thermal equilibrium in the early universe. In principle, all three Sakharov conditions could be satisfied in the context of the SM: strong electroweak (EW) or quantum chromodynamics (QCD) PTs could provide the necessary departure from thermal equilibrium, the chirality of the EW sector maximally breaks $C$, the Cabbibo-Kobayashi-Maskawa (CKM) matrix responsible for the quark mixing includes a $CP$-violating phase and sphaleron interactions violate the conservation of baryon number.
However, in the context of the SM neither the QCD~\cite{Aoki:2006we} nor the EW~\cite{Kajantie:1996mn} PTs are first order, and so the departure from thermal equilibrium is not enough to explain the observed value of BAU. Moreover, the CP-violation contained in the CKM matrix is below the amount needed to account for the observed BAU~\cite{Gavela:1993ts}. 
Hence, it is tempting to conclude that explaining the value in  Eq.\eqref{eq:observed_BAU} requires  BSM physics.

An attractive way to explain the observed BAU in the context of BSM, is to first produce an asymmetry in the lepton number and then transfer it to the baryons via sphaleron interactions. This is realized through the scenario of \emph{leptogenesis}~\cite{FUKUGITA198645} which takes advantage of the out-of-equilibrium CP-violating decay of the right-handed neutrino to the SM. Several channels of production of the heavy fermion $N$, before their decay, have been explored in the literature, including the prototypical thermal leptogenesis~\cite{FUKUGITA198645}, the production via the decay of heavier particles~\cite{Cataldi:2024bcs, Tong:2024lmi, Dick:1999je} or via a FOPT~\cite{Katz:2016adq,Azatov:2021irb, Chun:2023ezg, Huang:2022vkf, Cataldi:2024pgt}. A somewhat particular feature of leptogenesis via FOPT is that since the right-handed neutrino can be produced non-thermally, its mass can be much heavier than the temperature of the ambient plasma when it decays. Consequently, leptogenesis via the BC mechanism can be achieved even if the reheating temperature of the Universe never reaches the mass scale of the right-handed neutrino. For the same reason, strong washout can also be avoided even for heavy values of the right-handed neutrino mass, $m_N \simeq 10^{14}$ GeV, as typically dictated by the type I seesaw mechanism. 

The possibility to decouple the mass of the right-handed neutrino from the temperature of the plasma has motivated several studies of leptogenesis via the \textit{decay} of the right-handed neutrino that is produced in the presence of ultrarelativistic PT bubbles. However, in the context of PTs, to the best of our knowledge, the CP- violation in the \emph{production} of the heavy fermions $N$ during BCs or in the \emph{direct production} of light SM states via off-shell $N$ : $\phi^\star \to \chi HL$ has not been investigated in depth (with the exception of \cite{Azatov:2021irb}, that considered the production mechanism from the BP interactions but not the BC mechanism.). This is what we aim to do in this section.

To begin, we generalise the Lagrangian in Eq.\eqref{eq:main_model} to include more than one flavour of $\chi$ and $N$
\bea 
\label{eq:Lag_CP_viol}
\mathcal{L} = \sum_{iI} Y_{iI} \phi  \bar N_i  P_R\chi_I +  \sum_{i\alpha} y_{i\alpha}P_R N_i (\tilde H \bar L_{\alpha}) + \sum_{i} m^N_i \bar N_i N_i+ \sum_{I} m^\chi_I \bar \chi_I \chi_I - V(\phi,T) + h.c.
\eea 
where the indices take the values $i,I=1,2$.

Notice that there are several terms allowed by the presently imposed symmetries that we have not included in our Lagrangian. The opposite chirality for the $\chi N \phi$ term is examined in Appendix \ref{app:chiral_2}, while the effect of coupling of $\chi$ to the SM is discussed at the end of this section. Furthermore, we have not included any $L$ violating interactions, since the masses of all of the gauge singlet fermions are of the Dirac nature, and the field undergoing the PT $\phi$ is \emph{not} charged under lepton number. We will discuss the inclusion of lepton-number-violating terms in section \ref{sec:seesaw}.

\subsection{CP-violation from bubble collisons}
\label{sec:asymmetry}

We now compute the CP-asymmetry resulting from the BC mechanism, separately considering the two-body processes $\phi^\star \to \chi^c N$ followed by $N \rightarrow H L$, and the three-body process $\phi^\star \to \chi^c HL$. 

\subsubsection{CP-violation in the production of heavy fermions}

We start with the production of a pair of heavy on-shell particles  via $ \phi^\star \to N_i \chi^{c}_I$, where we use capital indices for $\chi$, lower case indices for $N$ and Greek letters for the SM families $L_\alpha$.\footnote{Notice that we do not use Einstein summation convention.} Using the model in Eq.\eqref{eq:Lag_CP_viol}, there are three distinct diagrams up to 1-loop level, which are illustrated in Fig.\ref{fig:CP_viold}. 
The CP-asymmetry in the $N_i$ and $\chi_I$ populations immediately after the collision of bubbles is given by
\begin{align}
  \epsilon_{iI} &\equiv  \frac{| \mathcal{M}_{\phi \to N_i \Bar{\chi}_I}|^2 -| \mathcal{M}_{\phi \to \Bar{N}_i \chi_I}|^2}{\sum_{iI} | \mathcal{M}_{\phi \to N_i \Bar{\chi}_I}|^2 +| \mathcal{M}_{\phi \to \Bar{N}_i \chi_I}|^2} \nonumber \\
 &=\frac{2\sum_{j,J} {\rm Im} (Y_{iI} Y_{iJ}^* Y_{jJ} Y_{jI}^* ) {\rm Im }f^{(\chi\phi)}_{ij}}{ \sum_{i,I}|Y_{i I}|^2}+\frac{2\sum_{\alpha,j} {\rm Im} (Y_{iI} y_{i \alpha }y_{j \alpha}^\ast Y_{jI}^*) {\rm Im}f^{(HL)}_{ij}}{ \sum_{i,I}|Y_{i I}|^2},
 \label{eq:CP_asym_1}
\end{align}
where $\epsilon_{iI}$ refers to asymmetry in the population of $N_i \chi_I^c$ with respect to that of $N_i^c \chi_I$. 
At this point, let us emphasize that the coupling to the SM is crucial to avoid the freedom to define a further $U(1)$ which would enforce the cancellation of the CP-violation (see, for example \cite{Heeck:2023soj}). This can be avoided if $N$ has two different decay channels\footnote{For example, consider decoupling the heavy fermions $N_i$ from the SM, by having a Lagrangian of the form
\bea 
\mathcal{L} = \sum_{iI} Y_{iI} \phi  \bar N_i  P_R\chi_I  + \sum_{I} m^N_i \bar N_i N_i+ \sum_{I} m^\chi_I \bar \chi_I \chi_I + h.c.
\eea 
Then one could define a $\chi$ number such that $\chi[\chi] = 1, \chi[N] = -1, \chi[\phi] = -2 $, this symmetry immediately enforces that $\Gamma[N \to \phi \chi] = \Gamma[\bar N \to \bar \phi \bar \chi]$. Thus we need two different decay channels for $N$, which are given by the coupling to the $\chi$ and $\phi$ and the SM particles.
}.  

In the limit $m_N \gg m_{\chi}, m_{\phi} \gg m_L, m_H$ the loop functions take the form\footnote{From now on, for the sake of clarity, we will drop the subscript $\chi$ and $N$ whenever they are not necessary, and so we will use $m_{i} \equiv m_{N,i}$ and $m_I = m_{\chi, I}$, where the capitalisation of the index will indicate if the mass designates the $\chi$ or the $N$ fermion.}:

\begin{align} 
f^{(H L)}_{ij }(x) &\equiv 2\int \frac{d^4p}{(2\pi)^4}\frac{m_{i} m_{j}}{(p^2 - i\epsilon)((p-p_{\chi})^2 -i\epsilon)(p_{N}^2-m_{j}^2-i \epsilon)} \frac{ \Tr \left[
\slashed{p}_\chi \slashed{p} P_R\right]}{\Tr \left[ \slashed{p}_N \slashed{p}_\chi P_L  \right]}
\\
f^{(\chi \phi)}_{ij}(x) &\equiv \int \frac{d^4p}{(2\pi)^4}\frac{1}{(p^2-  i\epsilon)((p-p_{\chi})^2 - i\epsilon)(p_{N}^2-m_{j}^2-i \epsilon)} \frac{\Tr \left[ \slashed{p}_N 
\slashed{p}_\chi \slashed{p}_N   \slashed{p} P_L\right]}{\Tr \left[ \slashed{p}_N \slashed{p}_\chi P_L  \right]} \, ,
\end{align}
where the upper index ($(HL)$ or $(\chi \phi)$) identifies the particles running in the loop, $p_\chi$, $p_N$ denote the momentum of the respective final states and $p$ denotes the momentum of $H$ or $\chi$ in the loop.

The extra factor of 2 in the $f^{(HL)}_{ij}(x)$ function comes from the $SU(2)$ symmetry of the SM doublet. 
The imaginary part of those loop functions can be easily computed:
\begin{align}
\label{eq:Im_part_f}
& \text{Im}[f^{(HL)}_{ij}(x)] =\frac{1}{16\pi} \frac{\sqrt{x_{ij}}}{1-x_{ij}}, \qquad \text{Im}[f^{(\chi \phi)}_{ij}(x)] = \frac{1}{32\pi} \frac{1}{1-x_{ij}}  \, , \qquad  \qquad  x_{ij} = \frac{m_{j}^2}{m_{i}^2} \, . 
\end{align}
 
The final effect of the CP-violating interaction is to produce an imbalance between $\bar N_i \chi_j$ and $N_i \bar \chi_j$. However, since no interaction is violating the lepton number, no net lepton number can be produced at this step. What we observe instead is  a \emph{separation} of the lepton number in the $\chi$ and the $N$ sector, respectively
\bea 
n_{N_i}- n_{\bar N_{i}}= n^i_{\Delta N_i} \approx \sum_I\epsilon^{iI} n_{N_i} \qquad n_{\chi_I}- n{\bar\chi_I}=  n_{\Delta \chi_I} \approx -\sum_i \epsilon^{iI} n_{\chi_{I}} \, ,
\label{eq:DeltaN_Deltachi_at_production}
\eea 
with $n_{N_i}$ and $n_{\chi_{I}}$ denoting the number densities of $N_i$ and $\chi_I$ from the BC mechanism and  $\epsilon^{iI}$ given by only the contribution from the $HL$ loop:

\bea 
\label{eq:CP_prod}
\epsilon^{iI}_{\rm prod} =\frac{1}{16\pi} \frac{\sum_{ j \neq i, \alpha} \text{Im}\big[ y_{i \alpha }y_{\alpha j}^\ast  Y_{iI} Y^\star_{jI} \big] \frac{m_{i}m_{j}}{m_{j}^2-m_{i
}^2}}{\sum_{i,I}|Y_{i I}|^2} \, .
\eea 
In particular, in the limit that $m_{I} \simeq \Tilde{m}_{\chi}$, where $\Tilde{m}_{\chi}$ is a typical mass of $m_{I}$, after summing over flavors, the $\chi \phi$ loop does not contribute in creating asymmetry in N.\footnote{We have  ${\rm Im}\sum_{I} \epsilon^{(\phi \chi)}_{iI} = 0$, where the superscript emphasizes that we consider the contribution from the $\chi, \phi$ loop.  However, one should remember that the sum over $I$ is performed after multiplying with $n_{N_i}$ that implicitly also depends on $m_{\chi_I}$ (see Eqs.\eqref{eq:DeltaN_Deltachi_at_production} and \eqref{eq:Y_CB_app}  ). Instead in the limit that $m_{I} = \Tilde{m}_{\chi}$ for all $I$, we have $n^i_{\Delta N} = n_{Ni}\sum_{I} \epsilon^{(\phi \chi)}_{iI} = 0$. }   

We have seen how the production of $\bar\chi$ and $N$ can entail lepton number violation in their separate populations. However, since we have not introduced any lepton number breaking terms, the total change in the lepton number should still be zero. Indeed, one can easily check that
\bea 
n_{\Delta N} + n_{\Delta \chi} = 0 \, .
\eea 
The key difference between N and $\chi$, however, is that only $N$ interacts with the SM particles and thus can decay to light SM states, transferring its asymmetry to the SM population. If the abundances in $N$ and $\chi$ were left as such, the efficient interactions between the $\chi$, and $N$ would equilibrate the lepton number. However, once produced, the $N$ will quickly decay to i) the SM via $N \to HL$ and transfer its lepton asymmetry to the SM or ii) to the dark sector via $N \to \phi \chi$, partially equilibrating the asymmetry. Since $\chi$ cannot directly decay into the SM, a net lepton asymmetry in the SM can thus be produced. Taking into account the decay to the dark sector, we obtain 
\bea 
n_{\Delta L_\alpha}\big|_{\rm prod}\equiv n_{L_\alpha}- n_{\bar L_{\alpha}} \big|_{\rm prod}\approx \sum_I\epsilon^{iI}_{\rm prod} n_{N_i} \text{Br}[N_i \to HL_\alpha] \, . 
\eea

In section \ref{sec:Model} we will discuss how such a scenario can be implemented in a concrete particle physics model, leading to a realization of leptogenesis, and we explore how such a model can be modified to also account for the observed amount of DM, inducing \emph{cogenesis}. \\ \\

\subsubsection{CP-violation in the decay of the heavy Dirac fermion}

In addition to CP-violation in the production, there will also be a contribution to CP-violation in the decay of the produced $N$ particles. In the decay of the heavy fermion $N$ to $HL$, only the loop of $\phi \chi$ leads to CP-violation\footnote{Analogously to  the production of 
$N$, in this case the contribution of the $HL$-loop to the $n_L - n_{\bar L}$ asymmetry cancels when summed over the final state leptons, assuming that $m_{L_\alpha} \ll m_N$. }. The CP-violation parameter $\epsilon^{i \alpha}$ from the decay of $N_i$ is defined by:
\bea
\epsilon^{i\alpha} \equiv  \frac{| \mathcal{M}_{ N_i \to  L_\alpha  \tilde H}|^2 -| \mathcal{M}_{ \bar N_i \to  \bar L_\alpha H}|^2}{\sum_{i\alpha} | \mathcal{M}_{ N_i \to  L_\alpha  \tilde H}|^2 + | \mathcal{M}_{ \bar N_i \to  \bar L_\alpha H}|^2} \, . 
\eea 
The contribution of the $\phi \chi$-loop mentioned above, then gives
\bea 
\label{eq:CP_viol_decay}
\epsilon^{i\alpha}_{\rm decay} =\frac{1}{32\pi} \frac{\sum_{I, j \neq i} \text{Im}\big[ y_{j \alpha }^*y_{\alpha i}  Y_{iI} Y^\star_{jI} \big] \frac{m_{j} m_{i}}{m_{j}^2-m_{i}^2}}{\sum_{i,\alpha}|y_{i \alpha }|^2} \, .
\eea 
The lepton number asymmetry resulting from the decay of $N$ is thus given by
\bea 
n_{\Delta L_\alpha}\big|_{\rm decay} \equiv n_{L_\alpha}- n_{\bar L_{\alpha}}\big|_{\rm decay} \approx \sum_i\epsilon^{i\alpha}_{\rm decay} n_{N_i} \text{Br}[N_i \to HL_\alpha] \, .
\eea

\subsubsection{Total asymmetry from the production and decay of on-shell heavy states}

Combining the lepton number asymmetry resulting from production and the decay of the heavy fermion $N$, the total lepton number in the SM sector produced during the BC is given by 

\begin{align}
\label{eq:lepto_numb_heavy}
\frac{n_{\Delta L_\alpha}}{s(T_{\rm reh})} &\approx \frac{1}{s(T_{\rm reh})}\bigg(\sum_{i}\epsilon^{i\alpha} n_{N_i} +  \sum_{Ii}\epsilon^{iI} n_{N_i}\bigg) \text{Br}[N_i \to HL_\alpha] \, 
\notag \\
&\approx 
\frac{1}{32\pi}\sum_{i,  j \neq i} \frac{n_{N_i}}{s(T_{\rm reh})} \frac{m_{i}m_{j}}{m_{j}^2-m_{i}^2}\bigg( \frac{\sum_{I} \text{Im}\big[ y_{j \alpha }^*y_{\alpha i}  Y_{iI} Y^\star_{jI} \big] }{\sum_{i,\beta}|y_{i \beta }|^2} + 
2 \frac{\sum_{  \beta} \text{Im}\big[ y_{i \beta }y_{\beta j}^\ast  Y_{jI}^\ast Y_{iI} \big]}{\sum_{i,I}|Y_{i I}|^2}  \bigg) \text{Br}[N_i \to HL_\alpha] \, ,
\end{align} 
where $n_{N_i}$ is the  abundance of the heavy fermion $N_i$ given by Eq.\eqref{eq:Y_CB_app}. 

As noted in the previous section, as soon as the on-shell production is allowed $2\gamma_w v \gg m_N$, the final abundance and thus also the lepton number asymmetry, depends only logarithmically on the boost factor $\gamma_w$. 
\begin{figure}[t!]
\centering
\includegraphics[scale=0.4]{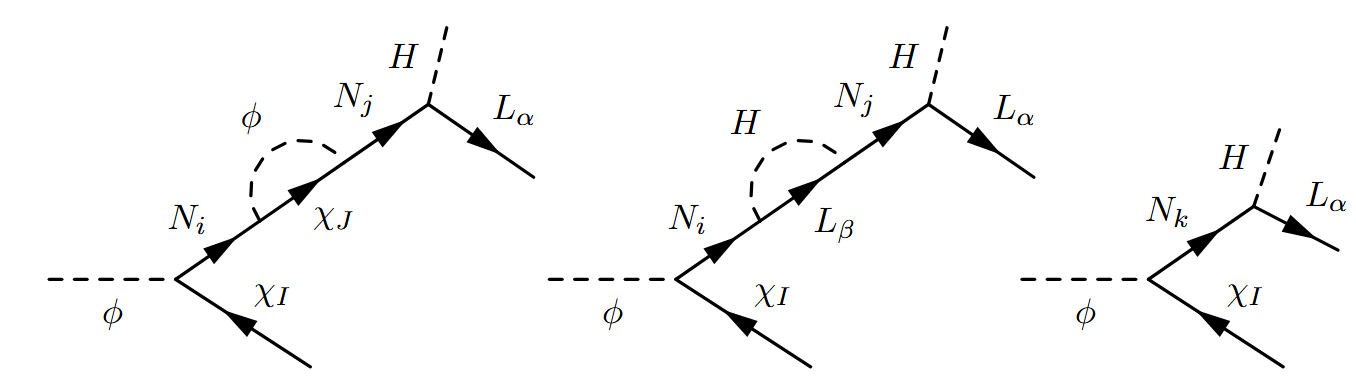}
\caption{ CP-violation resulting from the process $\phi^\star \rightarrow H L_\alpha \chi_I^c$.}
\label{fig:CP_viold_1_3}
\end{figure}

Another contribution to the production of baryon number comes from the collision of $\chi$ in the plasma and the wall condensate~\cite{Azatov:2020ufh, Azatov:2021irb}. This contribution is only sizable if $m_\chi \lesssim v$, i.e. the $\chi$ are abundant in the plasma and the mixing between $N$ and $\chi$ is large enough. The baryon number produced by the  BP interactions is of the order 
\begin{align}
\frac{n^{\rm BP}_{\Delta L}}{s(T_{\rm reh})}  &\simeq  -4 \times 10^{-5} \sum_{ij}\bigg(\frac{Y_iv}{m_i} \bigg)^2\frac{m_im_j}{m_i^2- m_j^2} \sum_{\alpha,J} {\rm Im} (Y_{Ii} Y_{Ij}^*y_{\alpha j}y^*_{\alpha i} )
\nonumber 
\\
&\times \l( \frac{2}{\sum_{iI}|Y_{ iI}|^2}-\frac{1}{\sum_{\alpha i} |y_{\alpha i}|^2}\r) \frac{1}{(1+\alpha)^{3/4}}
\text{Br}[N \to HL] \times  \frac{n^{}_{\chi}}{n^{\rm rel}_{\chi}} \times \theta(\gamma_w - m_i^2/(T_{\rm nuc}v)).
\end{align}
In the regime $m_\chi \gg T_{\rm nuc}$ we have $\frac{n^{}_{\chi}}{n^{\rm rel}_{\chi}} \sim e^{-m_\chi/T_{\rm nuc}}$. 
For $y \sim Y \sim {\rm Im}[y]\sim {\rm Im}[Y]$ and $m_2 \gg m_1$, this simplifies to 
\begin{align}
\label{eq:WP_partprod}
\frac{n^{\rm BP}_{\Delta L}}{s(T_{\rm reh})}  &\simeq  4 \times 10^{-5} \frac{Y^4 v^2}{m_2m_1}  \frac{\text{Br}[N \to HL]}{(1+\alpha)^{3/4}}
 \times  \frac{n^{}_{\chi}}{n^{\rm rel}_{\chi}} \times \theta\bigg(\gamma_w - \frac{m_1^2}{T_{\rm nuc}v}\bigg)\, . 
\end{align}
 We will compare this typical yield from the BP interactions with the one from the BC mechanism in Section \ref{subsec:paramspace}.

\subsection{CP-violation in the production of light SM particles}

Previously, we considered the CP-violation associated with the \emph{production} and \emph{decay} of heavy on-shell fermions. In addition, one can also consider the emission of $\phi^\star \to HL_\alpha \chi_I$ with the heavy fermion $N$ appearing as an off-shell intermediate state. Perhaps surprisingly, such channels also carry CP-violation, as we will describe now. The three diagrams contributing to the CP-violation are illustrated in Fig.\ref{fig:CP_viold_1_3}.

The contribution to the CP-violation is obtained by taking the imaginary part of the loop function, as done before. Again, we define the $\epsilon$ - parameter
\bea 
\epsilon \equiv \frac{|\mathcal{M}|^2_{\phi \to \Bar{\chi} \Tilde{H} L}- | \mathcal{M}|^2_{\phi \to \chi H\Bar{L}}}{|\mathcal{M}|^2_{\phi \to \Bar{\chi} \Tilde{H}L}+ | \mathcal{M}|^2_{\phi \to \chi H\Bar{L}}} \, . 
\eea 
For clarity, we split this parameter in terms of outgoing families, the fermion species in the propagators (denoted by subscripts) and the loop inserted in the diagram  (denoted by superscripts). Considering a loop of $HL$, one obtains
\begin{equation}
   \epsilon_{\alpha, \beta, I,i,j,k}^{(HL)} = -\frac{\Im \left[ y_{k \alpha}Y_{k I}y_{j \alpha}^\ast y_{\beta j}y_{\beta i}^\ast Y_{i I}^\ast\right]}{\sum_{l,r, I} \left(y_{r \alpha} Y_{r I}\frac{m_r}{(p_N^2 - m_r^2)}\right) \left(y_{l \alpha} Y_{l I}\frac{m_l}{(p_N^2 - m_l^2)}\right)^\ast}  \frac{p_N^2 m_i m_k}{(p_N^2 - m_i^2)(p_N^2 - m_j^2) (p_N^2 - m_k^2)} \, , 
\end{equation}
where $i,j$ and $k$ denote the generation of the heavy fermion N appearing in the loop and tree level diagrams respectively, $\alpha$ and $I$ denote the indices of the outgoing $L$ and $\bar \chi$, and $\beta$ denotes the family of the SM lepton running in the loop.

Obviously, this combination of couplings has a vanishing imaginary part if $i = j = k $. Assuming two generations for  $N$,  this leaves us with the cases $i \neq j = k$ and $k = i \neq j$.\footnote{The case $i=j$, corresponds to a loop correction to the mass of  $N$ in the propagator and thus does contribute to the CP-violation.} Summing over all intermediate states as well as over the outgoing $\bar \chi$ we then obtain for the $HL$-loop:

\begin{equation}
\begin{split}
\epsilon_{\alpha}^{(HL)} &= 
-\frac{2}{16 \pi } \sum_{i\neq j}\frac{\sum_{\beta, I}  \bigg( \frac{ \text{Im}\left[|Y_{i I}|^2 y_{i \alpha} y_{j \alpha}^\ast y_{\beta j} y_{\beta i}^\ast \right] m_i^2 p_N^2}{(p_N^2 - m_i^2)^2(p_N^2 - m_j^2)} + \frac{ \text{Im}\left[\abs{y_{j \alpha}}^2 Y_{jI}Y_{iI}^\ast y_{\beta j}y_{\beta i}^\ast\right] m_j m_i p_N^2}{(p_N^2 - m_i^2)(p_N^2 - m_j^2)^2} \bigg) }{\sum_{k,l, I} \left(y_{k \alpha} Y_{k I}\frac{m_k}{(p_N^2 - m_k^2)}\right) \left(y_{l \alpha} Y_{l I}\frac{m_l}{(p_N^2 - m_l^2)}\right)^\ast} \, . 
\end{split}
\end{equation}

This expression can be simply tested by scrutinizing its limit for $p_N \to m_i$ and $p_N \to m_j$. We obtain the two following limits

\begin{align}
\label{eq:HL_on_shell}
\epsilon_{\alpha}^{(HL)}(p_N \to m_j) &= 
\frac{2}{16 \pi \sum_{I}|Y_{Ij}|^2} \sum_{j\neq i}\sum_{\beta, I}  \text{Im}[ Y_{j I} Y_{I i}^{\ast} y_{i \beta}^\ast y_{j \beta} ] \frac{ m_j m_i}{(m_i^2 - m_j^2)}  \quad \text{CP-violating prod of $N_j$}
\\
\epsilon_{\alpha}^{(HL)}(p_N \to m_i) &= \frac{2}{16 \pi |y_{i \alpha}^2|}
\sum_{i\neq j}\sum_{\beta}  \text{Im}[ y_{i \alpha} y_{j \beta} y_{\alpha j}^\ast y_{\beta i}^\ast] \frac{  m_i^2}{(m_j^2 - m_i^2)} \quad \text{CP-violating decay of $N_i$}.
\end{align}

Those two pieces have a nice physical interpretation in terms of disconnected diagrams. $\epsilon_{\alpha}^{(HL)}(p_N \to m_j)$ can be interpreted as the production of an \emph{on-shell} $N_j$ in CP-violating way with an inserted $HL$ loop. In this sense, it corresponds to CP-violation counted in Eq.\eqref{eq:CP_prod}. On the other hand, the term $\epsilon_{\alpha}^{(HL)}(p_N \to m_i)$ is the CP-violating decay of the $N_i$ with an $HL$ loop. This term however, cancels after summing over $\alpha$, as was shown in the previous section.

Conversely, the CP-violation from the $\phi \chi$ loop reads
\begin{equation}
   \epsilon_{\alpha, \beta, I,i,j,k}^{(\chi \phi)} = -\frac{1}{16\pi}\frac{\Im \left[y_{k \alpha} Y_{k I} y_{\alpha j}^\ast Y_{jJ}^\ast Y_{Ji} Y_{I i}^\ast\right]}{\sum_{l,r, I} \left(y_{r \alpha} Y_{r I}\frac{m_r}{(p_N^2 - m_r^2)}\right) \left(y_{l \alpha} Y_{l I}\frac{m_l}{(p_N^2 - m_l^2)}\right)^\ast}  \frac{p_N^2 m_j m_k}{(p_N^2 - m_i^2)(p_N^2 - m_j^2) (p_N^2 - m_k^2)}
\end{equation}

Again, if we consider only two generations of $N$, then we have either $j=k$ or $i=k$.
Summing over the intermediate states and the outgoing $\chi$, we then have 

\begin{equation}
\begin{split}
\epsilon_{\alpha}^{(\phi\chi)} &= -\frac{1}{16 \pi} \sum_{i\neq j}\frac{\sum_{ I,J}  \bigg(\frac{m_i m_j p_N^2 \abs{Y_{Ii}}^2\text{Im}[ y_{\alpha i} y_{\alpha j}^\ast Y_{jJ}^\ast Y_{iJ} ]  }{(p_N^2 - m_j^2)(p_N^2 - m_i^2)^2} + \frac{m_j^2 p_N^2 \abs{y_{\alpha j}}^2 \text{Im}[ Y_{j I} Y_{I i}^\ast Y_{jJ }^\ast Y_{Ji}] }{(p_N^2 - m_j^2)^2(p_N^2 - m_i^2)} \bigg) }{\sum_{k,l, I} \left(y_{k \alpha} Y_{k I}\frac{m_k}{(p_N^2 - m_k^2)}\right) \left(y_{l \alpha} Y_{l I}\frac{m_l}{(p_N^2 - m_l^2)}\right)^\ast} \, . 
\end{split}
\end{equation}
Doing the same test as above by taking $p_N \to m_i, m_j$, we obtain

\begin{align}
\label{eq:epsilon_on_shell}
\epsilon_{\alpha}^{(\phi \chi)}(p_N \to m_i) &= \frac{1}{16 \pi  |y_{i \alpha}^2|}
\sum_{j\neq i} \sum_{ I, J}\frac{m_i m_j \text{Im}[  Y_{J i} Y_{jJ }^\ast y_{\alpha j}^\ast y_{\alpha i} ] }{(m_j^2 - m_i^2)} \quad \text{CP-violating decay of $N_i$}\\
\epsilon_{\alpha}^{(\phi \chi)}(p_N \to m_j) &= 
\frac{1}{16 \pi \sum_{I}|Y_{Ij}|^2} \sum_{i\neq j}\sum_{ I, J}   \frac{m_i m_j \text{Im}[Y_{j I} Y_{Ji} Y_{I i}^\ast Y_{j J}^\ast]  }{(m_i^2 - m_j^2)}  \quad \text{CP-violating prod of $N_j$}.
\end{align}

The piece $\epsilon_{\alpha}^{(\phi \chi)}(p_N \to m_j)$ describes the CP-violating production of $N_j$ and vanishes upon summation of final states as we discussed before. On the other hand, $\epsilon_{\alpha}^{(\phi \chi)}(p_N \to m_i)$ is the CP-violating decay of the $N_i$, already captured in Eq.\eqref{eq:CP_viol_decay}. 

The regime of \emph{resonant} leptogenesis $m_1 \to m_2$ implies resonance in the CP-violating piece. We however do not explicitly deal with this complication and assume $\Gamma_{1,2} \ll m_1-m_2$. We emphasize that our result might have relevant consequences on the scenario of leptogenesis catalyzed by the decay of a heavy particle like the inflaton~\cite{Asaka:1999yd, Hahn-Woernle:2008tsk, Barman:2021tgt} or heavy ALP~\cite{Cataldi:2024bcs}. 

As we have seen, the number of $H,L$ and $\chi$ can be computed in the following way 
\bea 
\frac{N_{\Delta L}}{A}\bigg|^{}_{\phi^\star \to \chi HL} 
\approx \int \frac{dp_zd\omega}{(2\pi)^2  } \Gamma^{\epsilon}_{\phi^\star \to \chi HL} (p) \big|\phi(p^2)\big|^2 \qquad \Rightarrow n_{\Delta L} \approx \frac{3 \beta H}{2 }\times  \frac{N_{\Delta L}}{A}\bigg|^{}_{\phi^\star \to \chi HL} \, ,
\label{eq:NoverA_3body}
\eea 
where $N_{\Delta L}$ designates the $N_L - N_{\bar L}$. Computing the lepton asymmetry requires including the CP-violation into the integral for the particle rate. The rate of the asymmetry production takes the form
\begin{equation}
\label{eq:1gto3CP}
\boxed{
\left(\Gamma^\epsilon_{\phi^\star \to HL \chi}\right)_{\alpha I} = \frac{1}{(2 \pi)^3} \frac{1}{32 p^3}  \int_{s_{12}^{\text{min}}}^{s_{12}^{\text{max}}} \int_{s_{23}^{\text{min}}}^{s_{23}^{\text{max}}} \abs{\mathcal{M}^{\alpha I}_{0}(s_{12}, s_{23})}^2 \epsilon_{\alpha I}(s_{12}) \text{d} s_{12} \, \text{d} s_{23}
}
\end{equation}
The $\epsilon$ - parameter is defined by
\begin{equation}
    \epsilon_{\alpha, I} = \sum_{i,j,k} \left( 
    \sum_{\beta} \epsilon^{(HL)}_{\alpha, \beta, I, i,j,k }
    + \sum_{J} \epsilon^{(\chi \phi)}_{\alpha, I,J, i,j,k}
    \right)
\end{equation}
and the tree-level spin-summed amplitude squared, including the resonance, is given by
\begin{equation}
    \abs{\mathcal{M}^{\alpha I}_{0}(s_{12}, s_{23})}^2 =  \sum_{q,r} y_{\alpha,q}Y_{q, I}y_{\alpha,r}^\ast Y_{r,I}^\ast \frac{m_q m_r (s_{23} - m_{L,\alpha}^2 - m_{\chi, I}^2)}{(s_{12} - m_q^2 + i m_q \Gamma_{N,q})(s_{12} - m_r^2 - i m_r \Gamma_{N,r})} \, . 
\end{equation}

\subsubsection{CP-violation mediated by an on-shell $N_1$}

For simplicity, we assume that during the BC the $N_2$ is never produced on-shell, i.e. $m_2 \gg \gamma_w v$. The expression in Eq.\eqref{eq:1gto3CP} in the range where only $m_1$ can be produced on-shell, namely when
\begin{equation}
m_{\chi} + m_L + m_H \ll m_1 < p < \gamma_w v \ll m_2 
\end{equation}
then simplifies in the following way: first, the CP-violating parameter becomes dominated by the peak, so that $\epsilon_{\alpha I}(s_{12}) \to \epsilon_{\alpha I}(s_{12}= m_1^2)$ and can be factorised out of the integral. This CP-violating parameter receives two contributions in Eq.\eqref{eq:epsilon_on_shell} and Eq.\eqref{eq:HL_on_shell}. On the other hand, the pure production part can be computed in a similar way as in Eq.\eqref{eq:factorisation}. Putting everything together, the lepton number produced simplifies to Eq.\eqref{eq:lepto_numb_heavy}, accounting for the lepton number produced in the production and the decay of $N$.

\subsubsection{CP-violation mediated by an off-shell $N_i$}

\begin{figure}[ht]
    \centering
    \subfigure{
    \includegraphics[scale=0.58]{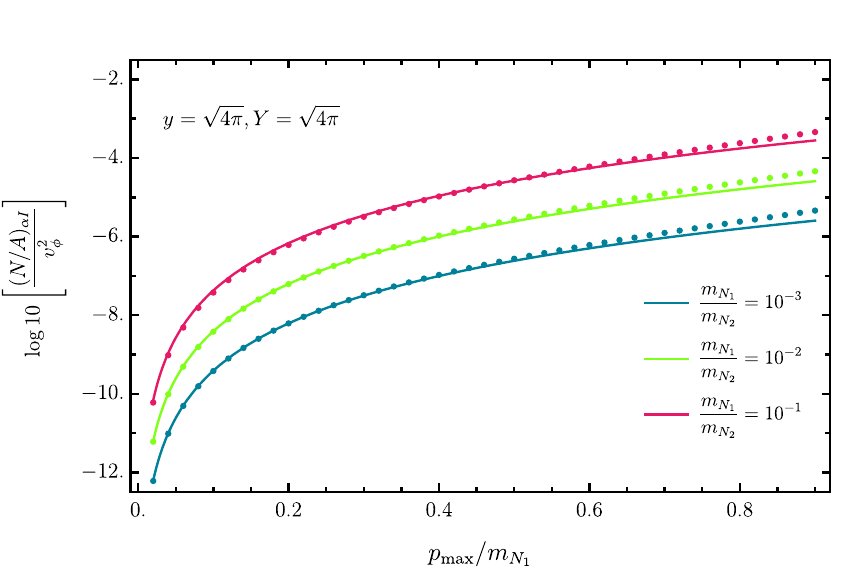}}
    \hspace{2mm}
    \subfigure{
    \includegraphics[scale=0.58]{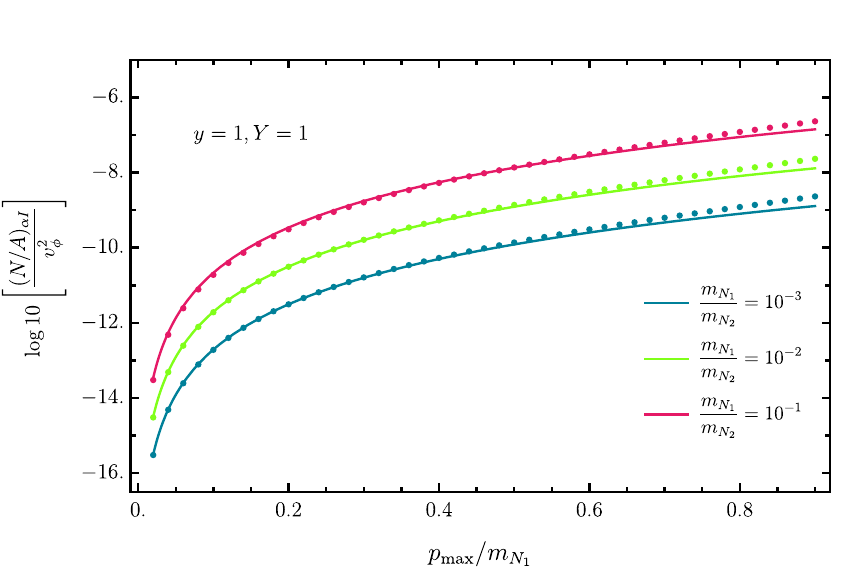}}
    \caption{Comparison of the off-shell production computed numerically (colored points) with analytic approximation (solid lines) using Eq.\eqref{eq:OFFshellAnalytic}, for $y=Y= \sqrt{4\pi}$ (left-panel) and $y=Y=1$ (right panel). }
    \label{Fig:NumVsAn}
\end{figure}

When the asymmetry is produced by purely off-shell N, $p_{\rm max} < m_{1,2}$, where $p_{\rm max}$ denotes the upper limit of the integral in Eq.\eqref{eq:NoverA_3body}, then for $m_2 \gg m_1$ the production rate can be approximated as
\bea 
\frac{N_{\Delta L
}}{A}\bigg|^{}_{\phi^\star \to \chi HL} \approx 10^{-3}\frac{  N \times C}{16\pi^6} \bigg(\abs{y}^4 \abs{Y}^2 \frac{m_1}{m_2} +  \frac{1}{2}\abs{y}^2 \abs{Y}^4\bigg) \frac{p_{\rm max}^4 v^2}{m_1^3m_2} \, , 
\label{eq:OFFshellAnalytic}
\eea 
in excellent agreement with numerics as displayed on Fig. \ref{Fig:NumVsAn}. Here $C$ is a factor counting the number of contributions with $ C \approx 3$, for the case where $Y=y$ and $C \approx 1/3$ for the case where $Y \ll y$. Since we mainly focus on the case $\abs{Y} = \abs{y}$ in our numerical studies of the parameter space, we will use $C \simeq 3$.

Putting everything together, we obtain for the lepton asymmetry in the SM $L_\alpha$, equal to the opposite lepton asymmetry in the dark sector $\chi_I$:
\bea 
\label{eq:asymm_offshell}
\frac{N_{\Delta L}}{A}\bigg|^{}_{\phi^\star \to \chi HL} 
\approx \int \frac{dp_zd\omega}{(2\pi)^2  } \Gamma^\epsilon(p)\big|\phi(p^2)\big|^2 \qquad \Rightarrow n_{L_\alpha} - n_{\bar L_\alpha} \approx \frac{3 \beta H }{2 (8 \pi)^{1/3}v_w }\times  \frac{N_{\Delta L}}{A}\bigg|^{}_{\phi \to \chi HL} \, ,
\eea 
where we called the asymmetry $\Delta L$. For low energies this expression can be computed numerically. 

\subsection{A possible suppression of the asymmetry}

Notice that we have not included another coupling, allowed by the present lepton number assignments, of the form: 
 \bea 
 \mathcal{L} \supset y_{\rm SM} \bar\chi (HL)  \, ,
 \eea 
which induces a partial decay $\chi$ to the SM instead of the dark sector made of $\tilde \phi, \tilde \chi$, that will be introduced in the next section. The presence of such a coupling would leave the relation in Eq.\eqref{eq_equal_sep} intact but could partially suppress the asymmetry. While the heavy N would decay partially to SM, successfully transmitting the asymmetry to the SM, a part of the carriers of the opposite charge, $\chi$, would also decay to the SM and cancel this newly created asymmetry. The final asymmetry would then be suppressed by the following branching ratio 
\bea 
Y_{\rm DS} = - Y_{\rm SM}, \qquad Y_{\rm SM} = Y_{\Delta L} \times \frac{\text{Br}[\chi \to \tilde \phi \tilde \chi]}{\text{Br}[\chi \to \tilde \phi \tilde \chi]+ \text{Br}[\chi \to HL]} \approx  Y_{\Delta L}\times \frac{y_1^2}{y_1^2 +y_{\rm SM}^2}\, .
\eea 
In this paper, we will assume that $\frac{y_1^2}{y_1^2 +y_{\rm SM}^2} \sim 1$, such that the effects of the $y_{\rm SM}$ can be neglected for simplicity.

\section{Implementation into a viable scenario}
\label{sec:Model}

 In the previous section, we have shown how the \textcolor{blue}{BC} mechanism can produce equal and opposite lepton number asymmetries in the SM and dark sector, 
\bea 
\label{eq_equal_sep}
Y_{\Delta L} = - Y_{\Delta \chi} \,  \qquad \text{(BC separation).}
\eea 

However,  equilibrating reactions between the SM and the dark sector could still erase the imbalance in the individual sectors. In this section, we  compute the equilibration rates of these reactions and the restrictions that are consequently imposed on the parameter space. 

Moreover, the lepton number stored in the dark sector could also play a role in determining the abundance of DM. This happens specifically if light dark sector particles with non-zero lepton number are coupled to $\chi$. In this case, when $\chi$ decays, it transfers its lepton number asymmetry to the light dark states,  providing a natural explanation to the \textit{coincidence problem} via the well-known mechanism of cogenesis.

Finally, we explore the implications of our model on light neutrino masses. We will see how Majorana masses for the heavy fermions can be induced by the PT, after $\phi$ gets a non-zero vev, allowing to explain the light neutrino masses via the seesaw mechanism.

\subsection{The danger of equilibration}

Since the $H,L$ particles are typically produced with very high initial energies, the qualitatively different \emph{thermal} and \emph{non-thermal} processes could both lead to the elimination of the lepton number. We incorporate all of the suppressions into a prefactor of the form 
\bea 
Y^{\rm fin}_{\Delta L} \approx \bigg(\Pi_i W_i\bigg)Y^{\rm init}_{\Delta L} \, ,
\eea 
where the individual contributions to wash-outs $W_i$ are computed in what follows.

\subsubsection{Non-thermal equilibration}

Particles emitted by the collisions of bubble walls are initially propagating in the plasma with very high energies. This includes specifically the  $HL$ and $\phi\chi$ pairs which carry the newly created lepton number.

A fast particle produced by the bubble wall collision, such as for example $L$, could subsequently scatter with a thermal $H$ from the plasma to produce an on-shell heavy $N$, equilibrating the two sectors.  To estimate this effect, we follow the reasoning of~\cite{Baldes:2021vyz}. The rate of production of an on-shell $N$ via this process can be estimated as~\cite{Baldes:2021vyz}
\bea 
\label{eq:inv_dec}
\Gamma_{HL \to N} \propto \frac{|y|^2}{8\pi} \frac{T m_N^2}{E_{L, \rm initial}^2} \times \text{Exp} \bigg[-\frac{m_N^2}{E_{L, \rm initial} T}\bigg] \, , 
\eea

where $E_{L, {\rm initial}}$ is the typical energy carried by the fast $H$ or $L$, which is initially of the order $\gamma_w T$. By inverting this relation, one can find the typical timescale for equilibration.

In thermal plasma, the fast $L$ thermalises mostly due to t-channel scatterings with gauge bosons with differential cross section of the form $d\sigma_{L W \to L W}/dt \sim g^4 (u^2+s^2)/64\pi t^2s^2$, and with energy exchange approximately given by $\delta E \sim -t/T$~\cite{Baldes:2022oev}. The time evolution of the energy of a fast $L$-particle is thus given by\footnote{Notice that elastic collisions  of non-thermal particles on the bath could lead to even faster thermalisation, suppressed however by the LPM effect~\cite{Harigaya:2014waa, Kurkela:2011ti}. We ignore those complications as they do not modify our qualitative conclusion.}
\bea 
\frac{dE_L}{d\tau} \approx n_W\int_0^{-s} dt \delta E \frac{d\sigma_{L W \to L W}}{dt} \approx -\frac{g^4T^2}{64\pi^3} \, .  
\eea 
By integrating over $\tau$, we find the typical thermalisation time:
\bea
\label{eq:thermtime}
\tau_{\rm therm} \sim \frac{64\pi^3}{g^4} \frac{E_{L, \rm initial}}{T^2} \, . 
\eea 
Comparing the typical  thermalisation time in Eq.\eqref{eq:thermtime} with the time of inverse decay computed in Eq.\eqref{eq:inv_dec}, it appears that even for larger energies $E_L$, the inverse decay timescale is always larger than the thermalisation timescale, and thus equilibration processes via non-thermal particles can be neglected. This can also be understood intuitively by noticing that unsuppressed production of $N$ via wall collision requires $\gamma_w T \gtrsim m_N$, while unsuppressed equilibration by scattering with the thermal plasma requires $\sqrt{\gamma_w} T \gtrsim m_N$, as we can see from the exponential in Eq.\eqref{eq:inv_dec}. 

\subsubsection{Equilibration via thermal scatterings with off-shell $N$: $W_{HL \to \phi \chi}$}

Thermal reactions of the form $HL \to \phi \chi$ mediated by the heavy fermion $N_i$ could also erase the lepton number from the two sectors. Since we are in the regime in which $T_{\rm reh} \sim v \ll m_N$,  we can compute the equilibration rate by integrating out the $N_i$ field.  For $E \ll m_N$, the squared amplitude averaged over initial spin states is given by 
\bea 
|\mathcal{M}_{HL \to \chi \phi }|^2 =\abs{y}^2 \abs{Y}^2\frac{m_\chi^2 - t}{m_N^2}, \qquad 
   \frac{d\sigma_{ HL \to \chi \phi }}{dt} =  \frac{1}{2}\frac{1}{\abs{\vec{p_L}}^2}\frac{|\mathcal{M}_{HL \to \chi \phi }|^2}{64\pi s}=\frac{|\mathcal{M}_{HL \to \chi \phi }|^2}{16\pi s^2},
\eea
where we used that in the center of mass frame, in the limit of zero masses for $L,H$, $\abs{\vec{p}_L} = \abs{\vec{p}_H} = \frac{\sqrt{s}}{2}$.
Assuming  that $m_H=m_L =0$ for simplicity, we integrate over t, with the integration limits given by
\begin{equation}
    t_{\rm min} = m_{\chi}^2 - \frac{s}{2} -\sqrt{s}\sqrt{\frac{s}{4} - m_\chi^2} \qquad t_{\rm max} = m_{\chi}^2 - \frac{s}{2} + \sqrt{s}\sqrt{\frac{s}{4} - m_\chi^2} \, , 
\end{equation}
and thus obtain
\begin{equation}
\sigma_{HL \rightarrow \chi \phi}(s) = \abs{y}^2 \abs{Y}^2\frac{\sqrt{1-\frac{4m_\chi^2}{s}}}{64 \pi m_N^2}.
\end{equation}
The \emph{thermally averaged} cross section $\gamma(ij \to kl)$  can be computed using~\cite{Davidson:2008bu}
\bea 
\gamma(ij \to kl) = \frac{g_i g_j T}{32\pi^4} \int ds s^{3/2} K_1(\sqrt{s}/T) \lambda\bigg(1, \frac{m_L^2}{s}, \frac{m_H^2}{s}\bigg) \sigma (s) \, ,
\eea 
where $\lambda(a,b,c) = (a-b-c)^2 - 4bc \approx 1 $, $K_1$ is a Bessel function of the second kind and $g_i$ are the internal degrees of freedom of the particle $i$. This expression has two useful limits, at $T \gg m_\chi$ and $T \ll m_\chi$ which are given by 
\begin{align}
\gamma^{T \gg m_\chi}(H L \to \chi \phi) &= \abs{y}^2 \abs{Y}^2 \frac{T^6}{64 \pi^5 m_N^2} \, ,
\\
\gamma^{T \ll m_\chi}(H L \to \chi \phi) &\approx \abs{y}^2 \abs{Y}^2 \frac{T^6}{32\pi^4} \times \frac{2}{64\pi m_N^2} \underbrace{\int^{\infty}_{m_\chi/T} dz z^4 \sqrt{\frac{\pi}{2z}} e^{-z} \bigg(1- \frac{4m_\chi^2}{z^2 T^2}\bigg)^{1/2}}_{\sim \frac{\pi}{\sqrt{2}} \left( \frac{m_\chi}{T}\right)^{7/2
} \rm{e}^{-2m_\chi/T}} \,  ,
\end{align}
where for $T \gg m_\chi$ we used that 
\bea
2T^5\int^{\infty}_{0} dx x^4 K_1(x) = 32T^5 \,  ,
\eea 
and for $T \ll m_\chi$ we used that
\bea
z K_1(z) \rightarrow \sqrt{\frac{\pi z}{2}} e^{-z}.
\eea

This leads to the following Boltzmann equation for the equilibration of lepton number
\bea 
H(x)x \frac{dY_{\Delta L}}{dx} \approx  - \gamma(HL \to \chi \phi) \frac{Y_{\Delta L}}{s Y_L} \quad \Rightarrow \frac{dY_{\Delta L}}{dx} \approx -\frac{M_{\rm Pl} x^4}{ 0.00031 g_\star^{3/2} m_\chi^5} \gamma(HL \to \chi \phi)(x) \times Y_{\Delta L}, 
\eea 
where $x = m_\chi/T$ and we have used $Y_L \approx 2.15 \times 10^{-3}$ for the relativistic abundance of the SM leptons. In order to estimate the maximum amount of wash-out, we solve the Boltzmann equation from reheating $x_{\rm reh} = m_\chi/T_{\rm reh}$  until the present time $x \rightarrow \infty$. In the relativistic regime $T_{\rm reh} \gtrsim m_\chi$, we obtain
\bea 
\label{eq:yield_equilibration_rel}
\frac{dY_{\Delta L}}{dx} \approx  -\abs{y}^2 \abs{Y}^2\frac{M_{\rm Pl} m_\chi}{0.3 \times 64\pi^5 m_N^2 x^2} Y_{\Delta L} \qquad \qquad \Rightarrow W_{HL \to \phi \chi} \approx \text{Exp}\bigg[- \frac{\abs{y}^2 \abs{Y}^2 M_{\rm Pl} T_{\rm reh}}{0.3 \times 64\pi^5 m_N^2 } \bigg] \, . 
\eea 
Instead, in the non-relativistic regime $T_{\rm reh} \ll m_\chi$, we obtain approximately the following evolution equation
\bea
\label{Eq:non_rel}
\frac{dY_{\Delta L}}{dx} \approx  -\abs{y}^2 \abs{Y}^2\frac{M_{\rm Pl} m_\chi}{445  \pi^{4} m_N^2} x^{3/2} e^{-2x} Y_{\Delta L} \, ,
\eea
which can be solved analytically to compute the value of $W_{HL \to \phi \chi}$ in the ultra non-relativistic case. We observe that in the deep non-relativistic regime, the washout process are almost completely switched off due to the Boltzmann suppression factor as one would expect. For the numerical studies of the parameter space that will follow we numerically interpolate between the analytic solutions for the extremely relativistic and nonrelativistic regimes. \\ \\

\subsubsection{Equilibration via thermal production of heavy $N$: $W_{H L \to N}$}

There is also equilibration from the direct production of $N$ via thermal $HL$ collisions. 
The Boltzmann equation which controls the evolution of the asymmetry is
\bea
s z H(z)\frac{d Y_{\Delta L}}{dz}=-\frac{Y_{\Delta L}}{Y_{L}}\gamma(HL\to N) \,,
\qquad   \qquad 
\gamma(HL\to N)=\frac{g_N T^3}{2\pi^2}z^2 K_1(z)\Gamma_{HL\to N} \, ,
\eea
where the Bessel function $K_1(z)$ with $z\equiv m_N/T$ has the two limiting behaviours
\bea 
zK_1(z) = \begin{cases}
1 \qquad z\ll 1,
\\
\sqrt{\frac{\pi z}{2}}e^{-z}\qquad z\gg 1.
\end{cases}
\label{eq_limit_cases}
\eea
Hence,  for large values of $z$, we get
\bea
\label{eq:condition2to1}
\frac{d Y_{\Delta L}}{d z}\simeq -\frac{2 e^{-z}z^{5/2}}{g_*^{1/2}}\l(\frac{M_{\rm pl}}{m_N}\r)\l(\frac{ \Gamma_{HL \to N}}{m_N}\r)Y_{\Delta L}, \qquad \Gamma_{HL \to N} \approx \l|\sum_{\alpha} y_{\alpha 1}
\r|^2\frac{m_N}{8\pi }.
\eea
Integrating this Boltzmann equation  from $z_{\rm reh} = m_N/T_{\rm reh}$ to $z \rightarrow \infty$, we obtain the wash-out factor
\bea 
\label{eq:equi_HL_N}
 W_{H L \to N}\approx  \text{Exp}\bigg[- \frac{2}{g_*^{1/2}}\l(\frac{M_{\rm pl}}{m_N}\r)\l(\frac{\l|\sum_{ \alpha} y_{\alpha 1}
\r|^2}{8\pi}\r) \times \Gamma\bigg[\frac{7}{2}, \frac{m_N}{T_{\rm reh}}\bigg]\bigg] \, ,
\eea 
where we only kept the contribution from the lighest $N$ species with mass $m_1$. 
Requiring that those production channels are suppressed amounts to ensuring that $m_N/T_{\rm reh} \gg 10$, implying a mild hierarchy between the scale of the PT and the mass scale of the heavy fermion $N_1$.

\subsection{Cogenesis realisation}
\label{sec:cogen}

\begin{figure}[ht]
    \centering
    \includegraphics[scale=0.5]{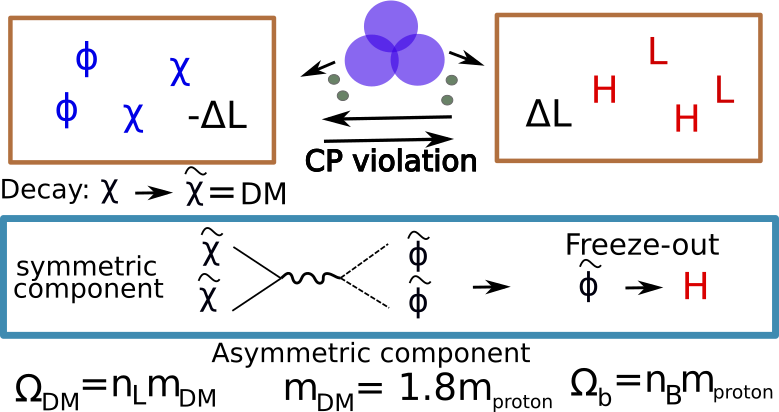}
    \caption{Scheme of the cascade of interactions leading to the production of ADM in the dark sector. The Lagrangian in Eq.\eqref{eq:model_DM} allows $\chi$ and $\phi$ to quickly decay into $\tilde \chi$ and $\tilde \phi$. The coupling of $\tilde\phi$ to $H$ will allow to freeze-out the abundance of $\tilde \phi$ and the symmetric abundance $\tilde \chi$. This requires typically large enough  couplings and $m_{\tilde \phi} \lesssim 500$ GeV.  }
    \label{fig:ADM}
\end{figure}

Now we slightly extend our previous scenario in such a way that the asymmetric dark sector abundance could account for the observed DM abundance. 
Due to the existence of asymmetry both in the SM and in the dark sector, a natural way to do this is by considering the scenario of asymmetric DM (ADM) (see \cite{Petraki:2013wwa, Davoudiasl:2012uw, Zurek:2013wia} for reviews and original papers). To that end, we extend the field content of the dark sector with a new fermion $\tilde \chi$, and a scalar $\tilde \phi$, both in the vector-like doublet representation under a new dark $SU(2)_D$ symmetry, which remains unbroken. Under the lepton number, those particles have assignments $L(\tilde \chi) =1, L(\tilde \phi) =0$. Let us consider the following extension of our former Lagrangian \cite{Borah:2024wos}
\begin{align}
\label{eq:model_DM}
    \mathcal{L}\supset y_{1}\bar \chi (\tilde \chi \tilde \phi)  +
     y_{2} \phi \bar \chi  \chi  
     + \lambda_{\tilde \phi H} |H|^2  |\tilde \phi|^2  - \frac{1}{2}m_{\tilde \chi} \tilde \chi \bar{\tilde \chi} - \frac{1}{2} m_{\Tilde \phi}^2 \abs{\Tilde \phi}^2 - \frac{\lambda_{\Tilde \phi}}{4} \abs{\Tilde \phi}^4  + 
     h.c.\, ,
\end{align}
where $y_{1,2}$ and $\lambda_{\tilde \phi H}$ are new dimensionless couplings.

In this case, all the asymmetry stored in the $\chi$ is transmitted to the $\tilde \chi$ via an interaction of the type $y_{1}\tilde \phi \bar \chi \tilde \chi$, allowing for the decay $\chi \to \tilde \phi \tilde \chi$. The $y_2 \phi \Bar \chi \chi$ interaction also makes the $\phi$ field unstable, and $\phi$ consequently decays to the $\tilde\phi$ and $ \Tilde \chi$ 4-body finally state via 2 off-shell $\chi$-s. In this minimal realisation, $\phi$ remnants might lead to a period of early matter domination, which can however, be avoided by complexifying the model. In this regard, we discuss in Appendix \ref{app:other_real} another model of the dark sector with similar properties, but free of any period of early matter domination.

Once an asymmetry has been produced in the dark sector, efficient interactions annihilate the symmetric component, leaving only the asymmetric one intact. The annihilation of the symmetric component of the dark sector proceeds via dark gauge-mediated $\tilde \chi \tilde \chi \to \tilde \phi \tilde \phi$ followed by the annihilation of $\tilde \phi$ to the SM via the Higgs portal. Efficient freeze-out of $\tilde\phi$ 
imposes that 
$m_{\tilde \phi} \lesssim 500$ GeV. 

Note that another  coupling of the form 
\bea 
 \mathcal{L} \supset \bar{\tilde\chi} (HL)  \, ,
 \eea 
which would induce a very efficient equilibration, is forbidden by the gauge symmetry imposed in the dark sector. $\lambda_{\phi \Tilde{\phi}} \abs{\Tilde \phi}^2 \abs{\phi}^2$ coupling could also be there in principle, but it has to be very small so that $\phi$ vev does not make $\Tilde \phi$ very heavy.

The energy fraction of the asymmetric component that remains unaffected by the annihilations depends on the mass of $\Tilde \chi$.
Requiring that its abundance explain that of DM, the DM mass is naturally set by the ratio of the DM abundance and the BAU,
\bea 
Y_{\Delta \tilde \chi} = Y_{\Delta \chi} = - Y_{\Delta L} = -  \frac{79}{28}  Y_{\Delta B}  \qquad \text{AND} \qquad \Omega_{\tilde \chi} \approx 5 \Omega_{b}  \qquad \Rightarrow \qquad m_{\tilde \chi} \approx 1.8 m_{\rm proton} \, , 
\eea
as is well-known in the models of ADM. Notice that the factor $\frac{28}{79}$ has been introduced to account for the sphaleron conversion rate.

\subsection{See-saw masses: light from heavy}
\label{sec:seesaw}

So far, we have not introduced any explicit breaking of the lepton number in the Lagrangian. One can however write the new terms

\bea 
\label{Eq:ToyMod}
\mathcal{L} \supset \sum_I\lambda_{N,R} \phi N_{R, I} \bar N_{R, I}^c + \lambda_{N,L} \phi N_{L, I} \bar N_{L, I}^c \, ,
%+ \lambda_{\chi, R}\phi \chi_{R, I} \bar \chi_{R, I}^c + \lambda_{\chi,L} \phi \chi_{L, I} \bar \chi_{L, I}^c \,. 
\eea 
which give Majorana masses to the $N$ particle after symmetry breaking. Notice that such interactions break the lepton number \emph{explicitly}, which is then not an accidental symmetry of the Lagrangian anymore. 
%, the mass matrix in the basis of the left-handed fields $\{\nu_L, N_L, N_R^c, \chi_L, \chi_R^c \}$ is given by
%\begin{equation}    M =      \begin{pmatrix}        0 && 0 && y  v_H && 0 && 0 \\ 0 && \lambda v && m_N && 0 && Y v \\ y^T v_H && m_N^T && \lambda v && 0 && 0 \\ 0 && 0 && 0 && \lambda v && m_\chi\\        0 && Y^T v && 0 && m_\chi^T && \lambda v    \end{pmatrix}    \label{eq:massmat}\end{equation}

Assuming that $\lambda_{N,L} = \lambda_{N,R} = \lambda$ and $(Yv)^2 \ll m_\chi^2$ for simplicity, the Lagrangian in Eq.\eqref{Eq:ToyMod} generates a dimension-5 Weinberg operator of the seesaw form~\cite{Minkowski:1977sc,Yanagida:1979as,GellMann:1980vs,Glashow:1979nm,Mohapatra:1979ia} 
\bea 
\mathcal{O}_{\rm Weinberg} = \sum_{I ,\alpha, \beta}\frac{y_{\alpha I}y^*_{\beta I}(\bar{L}^c_\alpha H)(L_\beta H ) \lambda v}{m_N^2}  \, , 
\eea
which induces a mass for the neutrinos 
\bea
\text{Max}[m_{\nu}] \sim \text{Max}\bigg[\sum_I|y_{\alpha I}|^2\bigg] \frac{  v_{EW}^2 \lambda v }{m_N^2}  \, . 
\label{eq:neutrino_mass}
\eea

Imposing that those masses recover the observed neutrino masses implies a constraint on our model, which is of the type 
\bea
\label{eq:lambda_val}
\lambda \sim 10^{-15} \frac{m_N^2}{\text{GeV} \times y^2_{\alpha I} v} \, .   
\eea 

This scenario  resembles the models of \emph{inverse} seesaw~\cite{PhysRevD.34.1642,Deppisch:2015qwa}. The perturbativity of the Yukawa coupling  $\lambda \lesssim \sqrt{4\pi}$ then implies
\bea
\label{eq:lambda_val_1}
\sqrt{4\pi}  \gtrsim 10^{-15} \frac{m_N^2}{\text{GeV} \times y^2_{\alpha I} v} \,  \qquad \qquad \text{(effective see-saw condition)} \, ,
\eea 
which is shown on the plot in Fig.\ref{fig:parspace_onshell} with a dotted-dashed line. Above this line, the condition in Eq.\eqref{eq:lambda_val_1} cannot be fulfilled and one needs to assume another unrelated mechanism for the production of neutrino masses. 

On top of this constraint, of course, adding the interactions in the Lagrangian Eq.\eqref{Eq:ToyMod} opens the possibility of producing two heavy fermions $N$ via $\phi^\star \to NN$, with subsequent CP-violating decays to SM, thus creating some additional lepton number. For the sake of simplicity, we neglect this additional contribution to the lepton number asymmetry in the present study.   

\subsubsection{Impact of wash-outs from lepton violating interactions}

The Majorana mass introduced for the $N$ implies that there are lepton-violating interactions that can be a further source of washout of the lepton asymmetry. The most important among those processes is the $H^cL \to HL^c$ interaction, which is governed by the following Boltzmann equation
\bea 
s z H[z]\frac{dY_{\Delta L}}{dz} = - 2 \frac{\gamma_{H^cL \to HL^c} }{Y_L^{\rm eq}}  Y_{\Delta L}  \qquad \Rightarrow \qquad    \frac{dY_{\Delta L}}{dz} \simeq - \frac{6}{g^{1/2}_\star }\frac{ M_{\rm Pl}}{m_N^2} z \frac{\gamma_{HL \to H^cL^c} }{s Y_L^{\rm eq}}  Y_{\Delta L}  \, . 
\eea 

In the limit $T \ll m_N$, the scattering rate simplifies to~\cite{Buchmuller:2002rq}
\bea 
\frac{\gamma_{HL \to H^cL^c} }{n_L^{\rm eq}} \equiv \Gamma_{H^cL \to HL^c} = \Gamma_{LL \to H^cH^c}\approx \frac{T^3}{4\pi^3}  \frac{\sum m_{\nu_i}^2}{v_{\rm EW}^4},
\eea 
which decouples for $T \ll T_{\rm dec} \approx 3 \times 10^{13}$ GeV. We thus obtain for the wash-out factor
\bea 
\label{eq:wash-out}
   \frac{dY_{\Delta L}}{dz} = - \frac{6z^{-2}}{2\pi^3 g^{1/2}_\star }    \frac{M_{\rm Pl}m_N\sum m_{\nu_i}^2}{v_{\rm EW}^4} \times  Y_{\Delta L}
   \qquad \quad \Rightarrow \quad  W_{H^c L \to HL^c} \approx  \text{Exp}\bigg[- \frac{6}{2\pi^3 g^{1/2}_\star }    \frac{M_{\rm Pl}T_{\rm reh}\sum m_{\nu_i}^2}{v_{\rm EW}^4}\bigg] \, .  
\eea

\subsection{Summary and study of the parameter space}
\label{subsec:paramspace}

Let us now summarize the different aspects of the model, which addresses the 1) baryon number of the universe, 2) DM abundance and 3) mass of the light neutrinos by the inverse seesaw mechanism. \\

\begin{figure}[h!]
\centering
\subfigure{
\includegraphics[scale=0.34]{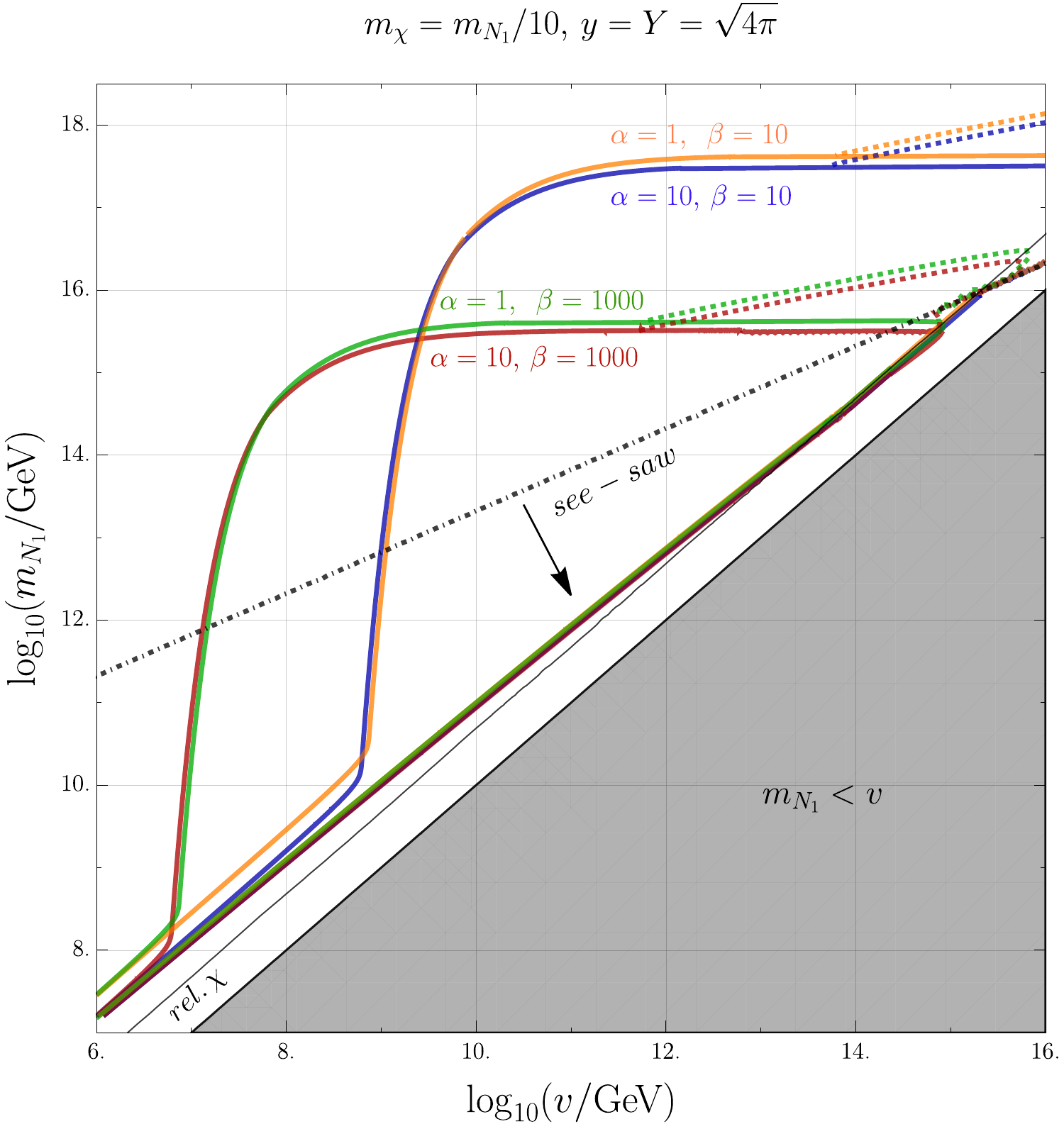}}
\hspace{2mm}
\subfigure{
\includegraphics[scale=0.34]{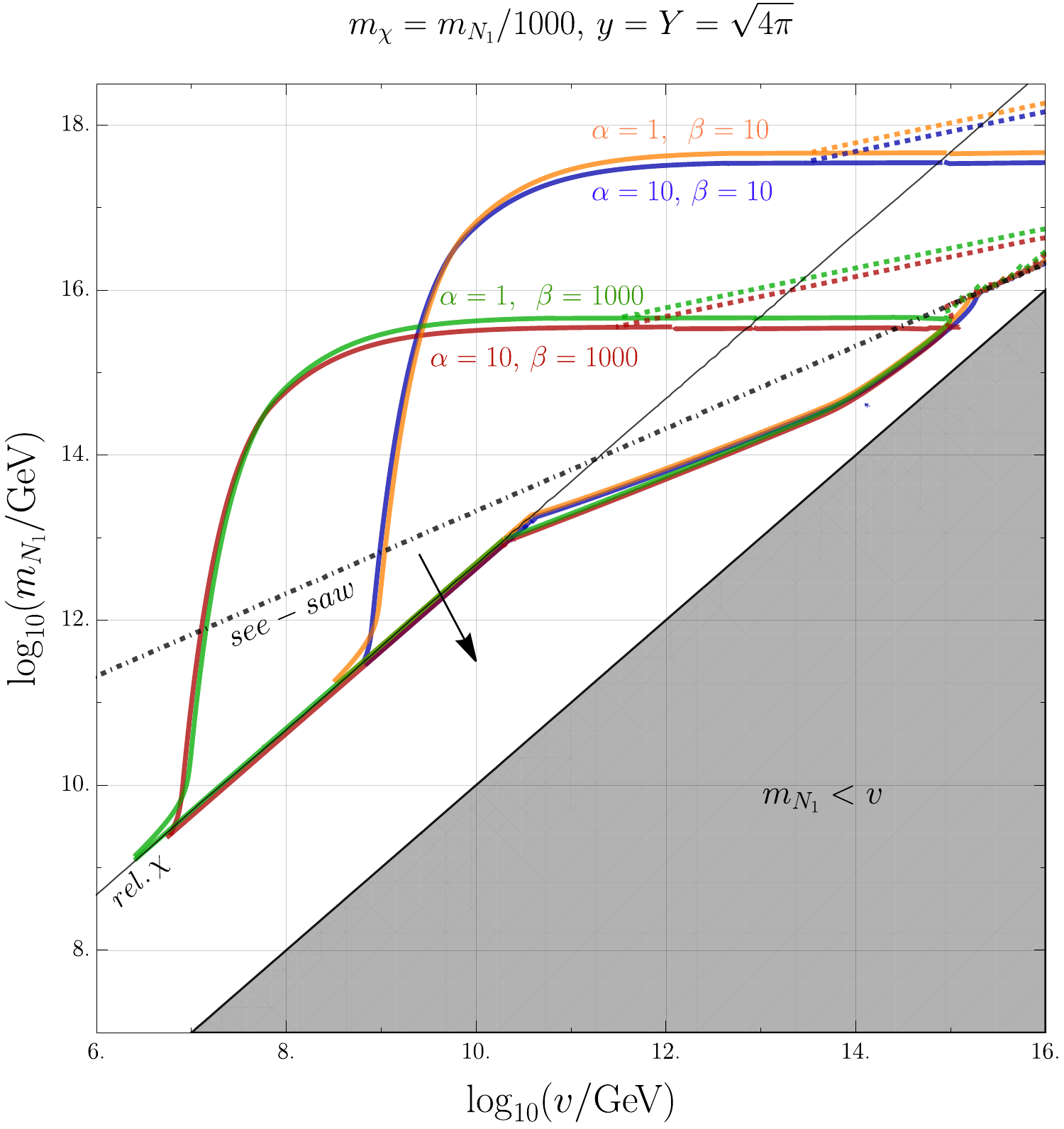}}
\hspace{2mm}
\subfigure{
\includegraphics[scale=0.34]{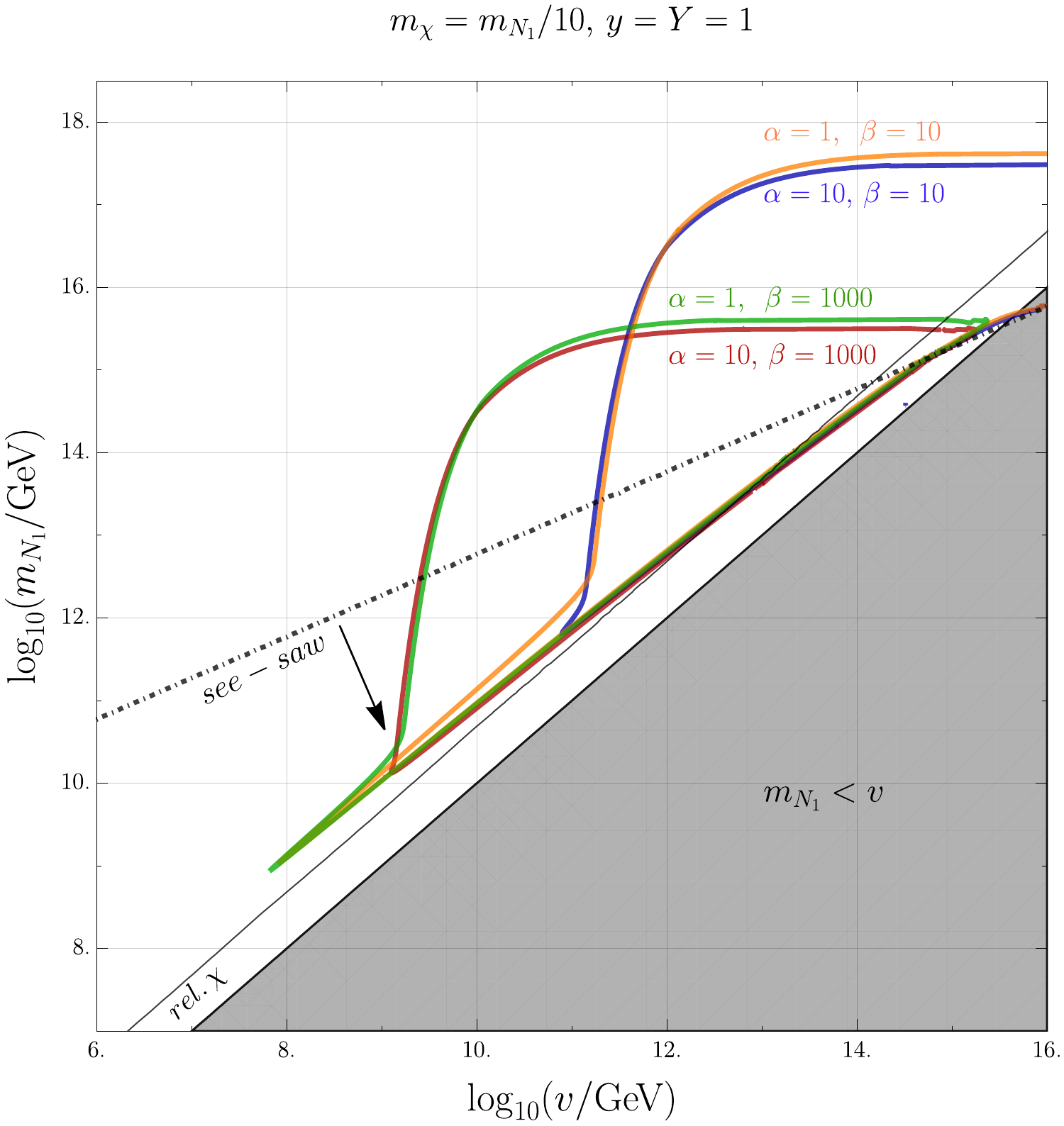}}
\hspace{2mm}
\subfigure{
\includegraphics[scale=0.34
]{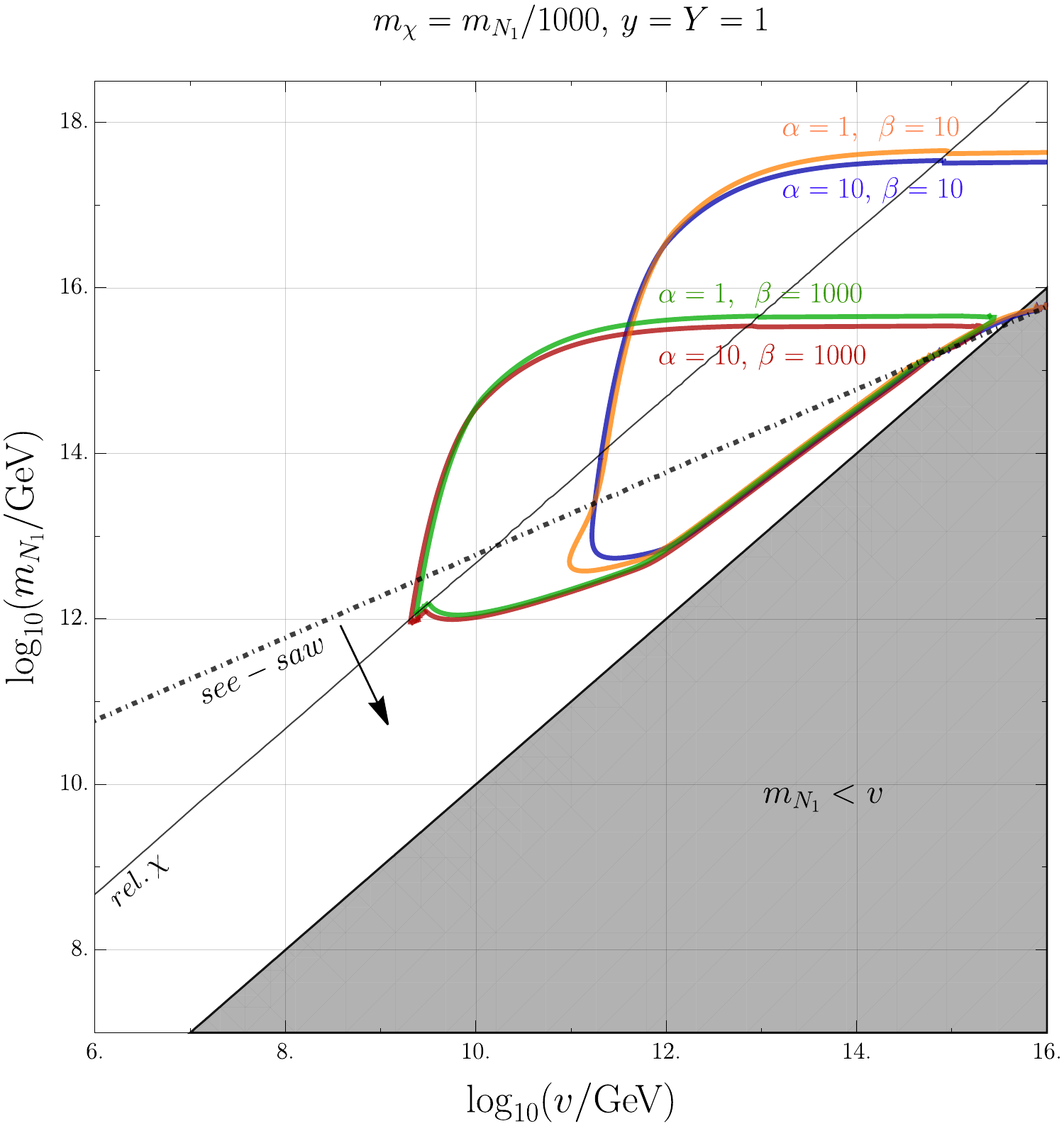}}
\caption{ \textbf{Top Left Panel}: Contours on the $v-m_{N_1}$ plane reproducing the observed BAU.  We set $m_{N_2}= 10 \, m_{N_1}$, $m_{\chi}= m_{N_1}/10$, $c_V=1$, $y= Y = \sqrt{4 \pi}$. The \emph{solid} part of the curves features the asymmetry produced by \emph{on-shell} production $\phi^\star \to \chi N$, while the dashed part of the lines proceeds via the \emph{off-shell} production $\phi^\star \to \chi HL$.  Above the ``see-saw'' dotted-dashed line, from the perturbativity condition in Eq.\eqref{eq:lambda_val_1}, the model cannot reproduce the light neutrino masses, while below the line it can. At $T= T_{\rm reh}$, the particles $\chi$ are relativistic below the solid black line named ``rel. $\chi$''.  \textbf{Top Right Panel}:  Same plot for $m_{\chi} = m_{N_1}/1000$ and  $y= Y=\sqrt{4 \pi}$.  \textbf{Bottom Left Panel}:  Same plot for $m_{\chi} = m_{N_1}/10$ and $y= Y=1$. \textbf{Bottom Right Panel}:  Same plot for $m_{\chi} = m_{N_1}/1000$ and $y= Y=1$.} 
\label{fig:parspace_onshell}
\end{figure}

The first important aspect is the effect of the different wash-outs and equilibration processes, which can be accounted by a simple multiplication
\bea 
\label{eq:fin_lept_asymm}
Y^{\rm fin}_{\Delta L} \approx \underbrace{W_{H^c L \to HL^c}}_{\text{Eq.}\eqref{eq:wash-out}}\times \underbrace{W_{H L \to N}}_{\text{Eq.}\eqref{eq:equi_HL_N}}\times\underbrace{W_{H L \to \phi \chi }}_{\text{Eq.\eqref{eq:yield_equilibration_rel} and Eq.\eqref{Eq:non_rel}}} \times \quad  Y^{\rm init}_{\Delta L} \, ,
\eea 
where the wash-out factors $W_i$ have been computed previously and encode the different sources of suppression. On Fig.\ref{fig:parspace_onshell}, we present the numerical study of the mechanism we discussed in this paper. The contours presented allow to recover the observed baryon abundance after sphaleron conversion via $ Y_{\Delta B} = \frac{28}{79} Y_{\Delta L}^{\rm fin}$ in the window $v \sim [10^7, 10^{16}]$ GeV, where the final lepton abundance $Y_{\Delta L}^{\rm fin}$ is given by Eq.\eqref{eq:fin_lept_asymm}. We present four benchmark scenarios with fixed parameters $m_{N_2}= 10\, m_{N_1}$, $c_V=1$ while varying $m_\chi$ and $y, Y$. Smaller Yukawa couplings, e.g. $y=Y=1$ (bottom panels in Fig.\ref{fig:parspace_onshell}), produce less lepton number carrying heavy fermions $N$ and, consequently, less baryon asymmetry, but allow to avoid backreactions (see Eq.\eqref{eq:backreact_cond}) in the whole parameter space, which partially dominate with $y=Y=\sqrt{4 \pi}$ (top panels), even though we expect that backreactions only affect the particle production by the BC mechanism logarithmically, as explained in section \ref{sect:UV_cutoff}. Instead, the value of the mass $m_{\chi}$ affects the impact of the equilibration processes $\phi \chi \ra H L$: a lighter $\chi$ can attain the relativistic regime in a larger portion of parameter space, thus washing-out the BAU produced. Below the solid black line the $\chi$s are relativistic, while above that they are not. Notice that for $v\gtrsim 3 \times 10^{13}$ GeV, the produced abundance is washed out immediately by the $HL^c \to H^c L$ process, which is only unavoidable if we require that our model is compatible with the inverse see-saw scenario, that is possible only below the dotted-dashed curve. Moreover, the black-shaded region where $m_{N_{1}} < v$  is excluded, since several assumptions of the BC particle production formalism do not hold there anymore.

In order to investigate the amount of BAU produced by the BC, we carefully distinguish between the \textit{on-shell} and \textit{off-shell} contributions, represented by the solid and dashed lines in Fig.\ref{fig:parspace_onshell}, respectively. The solutions for the \textit{on-shell} $N$ production via the decay $\phi^\ast \ra \chi N$, i.e. when $m_{N,1} \ll \gamma_w v$,  are computed using Eq.\eqref{eq:lepto_numb_heavy}.  In this case, there are typically two solutions for $m_{N_1}$ for a given value of $v$ - at small and large $m_{N_1}$ - determining thus two branches in $m_{N_1}$: the upper one where the curves for different $\beta$ and $\alpha$ are split and the lower one where those curves merge. One can observe that the lower branch is controlled by the equilibration rates $\phi \chi \to HL$, while the upper branch is due to the suppression of the production by large $m_{N_1}$. Imposing the perturbativity condition for the see-saw masses in Eq.\eqref{eq:lambda_val_1} splits the parameter space into a region in which the light neutrino masses can be explained (below the dotted-dashed line) and a region in which they cannot (above the dotted-dashed line). Finally, let us stress that in our numerical studies we also implement the production of the heavy fermion N from BP interactions, contributing to the baryon asymmetry via Eq.\eqref{eq:WP_partprod} in parts of the parameter space. In particular, production from the BP interactions dominates over the BC mechanism production in the lower-left corner of the plots, inducing a spike in the solution $Y_{\Delta B}^{\rm fin}$. 

The dashed lines denote the \textit{off-shell} contribution, i.e. when $\gamma_w v \ll m_{N_1}$, where the production is dominated by the process $\phi^* \ra \chi L H$. The lepton asymmetry produced is thus given by Eq.\eqref{eq:asymm_offshell}. This extends the BC mechanism to even higher masses of $m_{N_1}$.

One can see that the produced baryon asymmetry has a mild dependence on $\alpha$, while increasing $\beta$ allows to reproduce the observed amount of BAU for smaller values of $v$ and $m_{N,1}$. To interpret this we notice that a larger $\beta$ reduces the boost factor, which decreases the yield of the BAU only logarithmically. On the other hand,  the increase of $\beta$ also implies a larger amount of bubbles per Hubble volume, i.e. more collisions, thus enhancing the particle number density and the BAU produced by the BC mechanism.

\subsection{Gravitational wave (GW) detectability}

In the previous sections we studied the production of BAU and DM as a consequence of PT dynamics. We saw that it required runaway walls or at least $\gamma_w \gg 1$, hinting at rather strong FOPTs. Such PTs are expected to induce a large background of GW due to the sound waves in the plasma and the collision of bubble walls, making this mechanism possibly detectable via GW interferometers. The GW signal from a FOPT is expected to resemble a broken power-law with peak frequency around 
\bea 
f_{\rm GW} \sim   \text{Hz} \times\beta \bigg(\frac{v}{10^8 \text{GeV}}\bigg) \, . 
\eea 
This suggests that part of the parameter space where the model we presented is successful overlaps with the sensitivity range of the future Einstein Telescope (ET) observatory. 

After the bubbles have collided, the energy of the PT is transmitted to very thin plasma and scalar gradient shells, which keep propagating in the Universe. The anisotropic stress sourced by those shells induces GWs. For $\gamma_w \gg 1$, the behaviour of such shells of plasma and scalar field is best described by the \emph{bulk flow} model \cite{Konstandin:2017sat}
\footnote{The authors thank Jorinde Van De Vis and Ryusuke Jinno for helpful discussions on the bulk flow model.} (see also \cite{Lewicki:2022pdb} for another model adequate for strong interactions). In the runaway or effectively runaway regime relevant to us, bubble wall motion is expected to produce extremely thin and highly relativistic fluid shells around its scalar profile, which evolve into long-lived shock waves following bubble collisions~\cite{Jinno:2019jhi}. The significant disparity in scales between the bubble radius and the shock front thickness presents a major challenge for numerical simulations. However, from the perspective of GW generation, a sharply peaked momentum distribution in the plasma is expected to be indistinguishable from the one carried by the scalar field. Consequently, the resulting gravitational wave signals in both cases should be similar and be described by the bulk flow model, which was initially designed to capture the GW signal from relativistic scalar shells. \cite{Gouttenoire:2023bqy, Baldes:2023cih}. Finally, a recent study\cite{Jinno:2022fom} conducted in the moderately relativistic regime ($\gamma_w \lesssim 10$) suggests that the GW spectrum once again resembles the one predicted by the bulk flow model. 
%%%%%%%%%%%%%%%%%%%

Following Ref.~\cite{Konstandin:2017sat} the GW signal, assuming $v_w \to 1$, takes the form
\bea 
h^2\Omega^{\rm today}_{\rm GW} = h^2\Omega_{\rm peak} S(f, f_{\rm peak}) \qquad 
S(f, f_{\rm peak}) = \frac{(a+b) f_{\rm peak}^b f^a}{b f_{\rm peak}^{(a+b)}+ a f^{(a+b)} }, \qquad (a, b) \approx (0.9, 2.1) \, , 
\eea 
with the energy density parameter $\Omega_{\rm peak}$ and the peak frequency $f_{\rm peak}$ given by 
\begin{align}
h^2\Omega_{\rm peak} &\approx 1.06 \times 10^{-6} \beta^{-2} \bigg(\frac{\alpha\kappa}{1+\alpha}\bigg)^2 \bigg(\frac{100}{g_\star}\bigg)^{1/3} \, ,
\nn
f_{\rm peak} &\approx 2.12 \times 10^{-3} \beta \bigg(\frac{T_{\rm reh}}{100 \text{GeV}}\bigg) \bigg(\frac{100}{g_\star}\bigg)^{-1/6} \quad \text{mHz} \, .
\end{align}
Note that on the top of this spectrum, one needs to impose an IR cut-off required by causality for $f < H_{\rm reh}/2\pi$, where  $H_{\rm reh}$ denotes the Hubble rate evaluated at the reheating temperature after the transition has completed. Since the energy of the transition goes mostly to the sound waves and to the scalar field, we can set $\kappa=1$ to capture those two contributions. 

Moreover, GWs could also be produced by free-streaming heavy particles produced by the BC \cite{Inomata:2024rkt} (see also \cite{Jinno:2022fom}). We do not consider this source in the current paper, because the heavy particles are strongly interacting and decay fast: the $N$ decays in a timescale $\tau^{\rm decay}_{N} \sim \frac{8\pi}{|y|^2}\frac{\gamma_w}{m_N} \sim \frac{8\pi}{|y|^2}\frac{M_{\rm Pl}}{\beta v m_N }$, which is parametrically smaller than the duration of the PT by a factor $v/m_N$. 

\begin{figure}[h!]
\centering
\subfigure{
\includegraphics[scale=0.455]{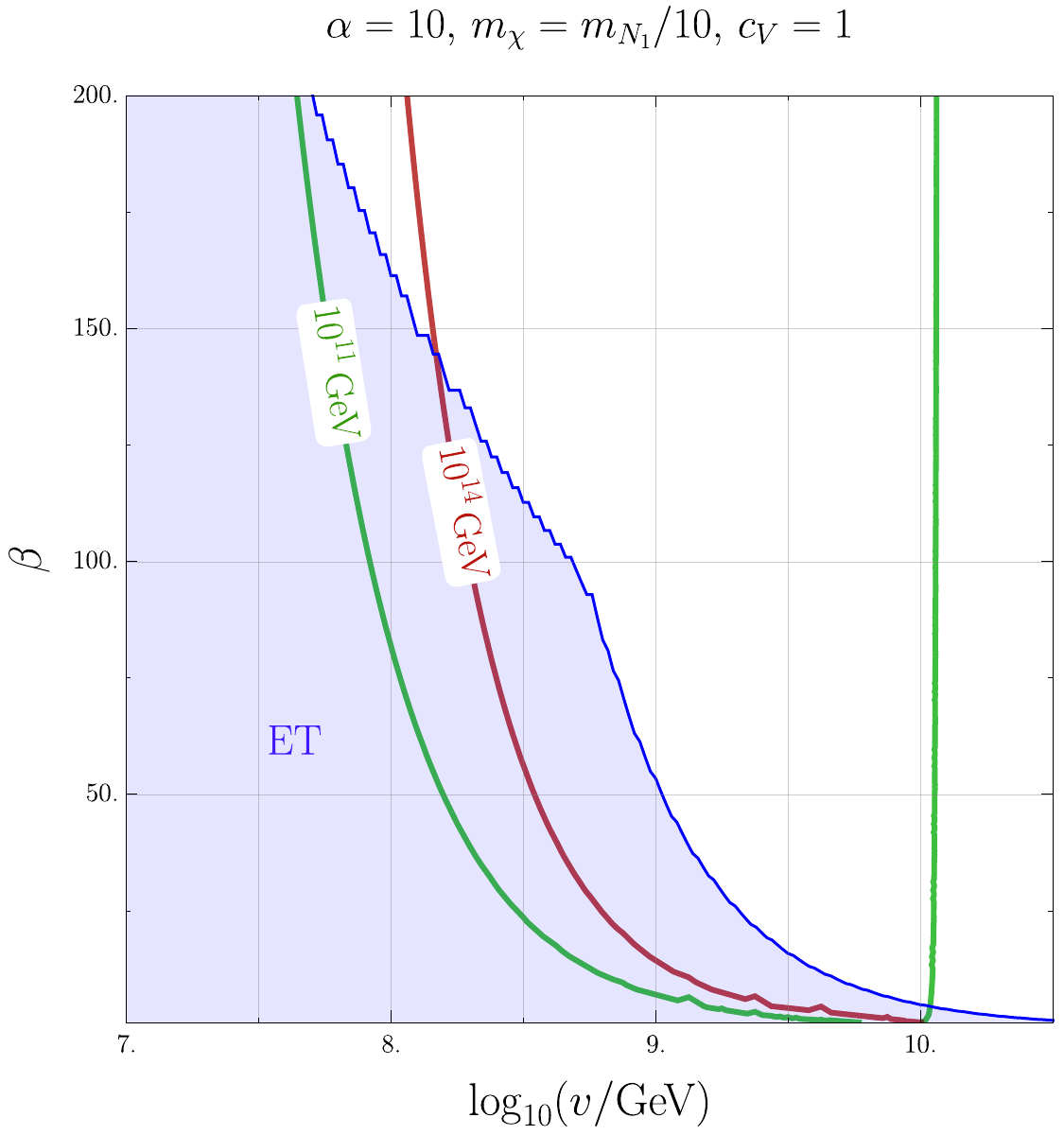}}
\hspace{1mm}
\subfigure{
\includegraphics[scale=0.455]{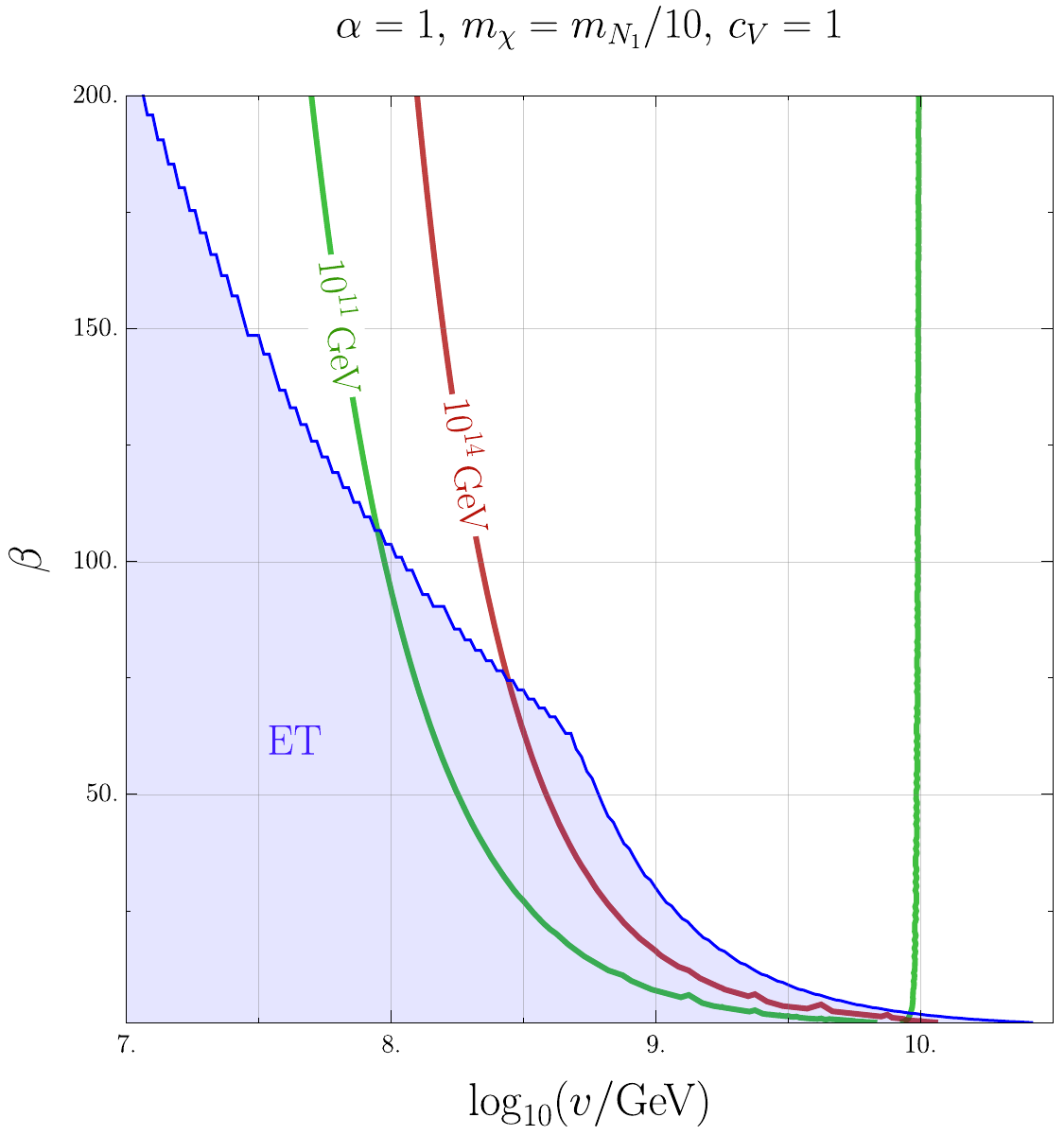}}
\caption{GW sensitivity curve for ET (blue-shaded region) and baryon asymmetry $Y_{\Delta B}$ produced in our scenario for $m_{N_1} = 10^{11}$ GeV (green line), $ 10^{14}$ GeV (red line), and $\alpha=10$ (left panel), $1$ (right panel). We set $y=Y= \sqrt{4\pi}$, $m_{\chi}= m_{N_1}/10, m_{N_2}= 10\, m_{N_1}$. The contour for ET is made requiring $\text{SNR}=1$~\cite{Moore:2014lga, Schmitz:2020syl} and ignoring the impact of an astrophysical foreground, which makes it a very optimistic contour. }
\label{fig:gw}
\end{figure}

We show the region of the parameter space that would be, in principle, observable at the future Einstein Telescope (ET) observatory~\cite{Moore:2014lga,Sathyaprakash:2012jk, Maggiore:2019uih} (see \cite{Abac:2025saz} for a review) on Fig.\ref{fig:gw}. We observe that models reproducing the observed baryon asymmetry can be detected at the ET observatory if the transition is slow enough, $\beta \lesssim 100$. As mentioned above, for a given value of the $m_N$ mass, there are typically two solutions for the scale of the symmetry breaking $v$, one corresponding to the left and upper branches of the triangle in Fig.\ref{fig:parspace_onshell}, controlled by the yield of fast particles, and one corresponding to the lower branch in Fig.\ref{fig:parspace_onshell}, controlled by the wash-outs. For a given $m_N$, the latter corresponds to a larger VEV than the former. We infer that only the regime which is controlled by the ``yield'' (and not by the wash-out) can be detected.

Let us conclude with a word of caution. The complicated problem of the separation of the unavoidable astrophysical background from the 
possible cosmological background is still under vivid investigation \cite{Caprini:2019pxz,Flauger:2020qyi,Boileau:2020rpg, Martinovic:2020hru}. This foreground is still subject to very large uncertainties and will depend on our abilities to resolve individual sources.   From the inspiral phase of the merger of black holes compact binaries, one expects a background of the form 
\bea 
\Omega^{\rm GW}_{\rm binaries} = \Omega_{\rm CBC}\bigg(\frac{f}{25 \text{Hz}}\bigg)^{2/3} \times \theta(f_{\rm cut} - f ) \, ,
\eea 
where $\Omega_{\rm CBC}$ is a constant that is expected to be extracted from observations. Its value is thought to be around $\Omega_{\rm CBC} \sim 10^{-9}$~\cite{LIGOScientific:2017zlf,Boileau:2020rpg}. The cut of the background comes from the merging of the lightest compact binaries, which we expect to be around a solar mass, corresponding to $f_{\rm cut} \sim 3 \times 10^3$ Hz. Indeed, the frequency associated to the innermost stable circular orbit (ISCO) when
the inspiral GW emission is close to maximal is $f_{\rm ISCO} = 4400 \text{Hz} M_{\rm sun}/(m_1+m_2)$, and so the merging of two solar mass black holes would lead to a maximal frequency around $f_{\rm cut}$. Motivated by the recent development of subtraction methods (see for example \cite{vanRemortel:2022fkb}), we assume in our analysis that this astrophysical background can be exactly removed, which is probably an optimistic assumption.

Our mechanism is also successful for PTs with GW signal peaking at frequencies higher than the ones observables at ET, in the range $f_{\rm peak} \gg 10^3 \, \text{Hz}$. To be properly explored, this range necessitates the development of new detectors, but has the virtue of being free from astrophysical GW background. Several proposals of detectors have already been put forward in such directions \cite{Aggarwal:2020olq}.   

\section{Conclusion}
\label{sec:conclusion}
In this paper, we have studied the production of heavy states from the collision of bubbles of a FOPT, and the associated CP-violation. In our setup, the FOPT \textit{does not break} the lepton number. We compute the CP-violation during 1) the production of $N$, via $\phi^\star \to \chi N$ and via 2) the direct production of light SM states $\phi^\star \to HL \chi$ and 3) during the decay to the SM $N \to HL$. Since the lepton number is not directly broken in  our setup, the lepton number in the visible sector is equal and opposite to the lepton number in the dark sector: the result of the BC is to separate the lepton number into the visible and the dark sectors. 

Through a direct implementation of our mechanism, we study a minimal model of cogenesis, where the negative lepton number in the dark sector cascades down to lighter particles with masses of the order of a few GeV. The BC mechanism is thus a natural process allowing for the separation of lepton number necessary for cogenesis, which explains both BAU and DM abundance via ADM with $m_{\rm DM} \sim 1.8 m_{\rm proton}$. We observe that such a cogenesis mechanism is compatible with the observations for a PT with the symmetry breaking scale in the range $v \sim [10^7, 10^{16}]$ GeV. For each given value of the VEV $v$, we observe that two values of $m_{N_1}$ can explain the observed abundance. 

To include the masses of the light neutrinos into the picture, we introduce an explicit lepton number violating coupling between the scalar field $\phi$ and the heavy fermion $N$ which produces a Majorana mass for the heavy fermion $N$ after the PT. The masses of the light neutrinos are then obtained via the familiar seesaw mechanism. We, however, observe that such a mechanism is only efficient to explain the light neutrino mass in a subset of the cogenesis parameter space.  However, in the current analysis, we have neglected the effect of the lepton number violating Yukawas in the production of asymmetry.

As a possible smoking gun of such a scenario, we discuss the GW signal emitted during the FOPT, modelling the GW signal using the bulk flow model. We observe that a subset of the parameter space, allowing cogenesis, could be detectable at the future GW observatory ET for $v \lesssim 10^{10} \text{ GeV} $. The BC mechanism thus offers a possibility to unify cogenesis with light neutrino masses, while being possibly detectable at future GW experiments.  

\section*{Acknowledgements}

It is a pleasure to thank Tomazs Dutka, Bibhushan Shakya and Enrico Nardi for insightful physics discussions and to thank Iason Baldes and Géraldine Servant for relevant comments on the manuscript. MV is supported by the ``Excellence of Science - EOS'' - be.h project n.30820817, and by the Strategic Research Program High-Energy Physics of the Vrije Universiteit Brussel. KM is supported by the Estonian Research Council personal grant PUTJD1256 and by the Estonian research council grants RVTT3, and the Center of Excellence program TK202 ``Fundamental Universe''. MC is supported by the Deutsche Forschungsgemeinschaft under Germany's Excellence Strategy - EXC 2121 ``Quantum Universe" - 390833306.

\appendix
\section{An alternative chirality assignment: the opposite chirality couplings}
\label{app:chiral_2}

In the main text, in Eq.\eqref{eq:main_model}, we opted for one particular chirality assignment. An alternative option is to consider 
\bea 
\label{eq:main_model_1}
\mathcal{L} =  Y \phi  P_L\bar N  \chi + \frac{1}{2} M_N N \bar N + \frac{1}{2}m_\chi \chi \bar \chi +  \sum_{\alpha} y_{\alpha}P_R N (\tilde H \bar L_{\alpha}) - V(\phi,T) \, . 
\eea 
where the only modification to the Lagrangian studied in the main text, Eq.\eqref{eq:main_model}, is the chirality assignment $Y \phi  P_L\bar N  \chi$. In this case, an opposite chirality of $N$ couples to $HL$ and $\phi\chi$ respectively, while in the main text it was the same chirality.  We however show now that this assignment leads to suppressed production of light particles and CP-violation, and so we do not study it further. 
\subsection{On-shell production}

For the the on-shell production, the chirality assignment obviously cannot impact the density of heavy particles produced. Consequently, Eq.\eqref{eq:prod_heavy_N} remains valid. 

\subsection{Off-shell production}
The rate of production is now given by 
\begin{equation}
 \Gamma_{\phi^\star \to HL \chi}(p^2) = \frac{2}{(2 \pi)^3} \frac{ \abs{y}^2 \abs{Y}^2}{32 \sqrt{p^6} } \int_{s_{12}^{\text{min}}}^{s_{12}^{\text{max}}} \int_{s_{23}^{\text{min}}}^{s_{23}^{\text{max}}} \frac{s_{12} (s_{23} - m_L^2 - m_{\chi}^2) }{(s_{12} - m_N^2)^2  + m_N^2 \Gamma_N^2} \text{d} s_{12} \, \text{d} s_{23}  \, , 
\end{equation}
where we notice that $s_{12}$ factor replaced the factor $m_N$ appearing in the main text. Since for the off-shell production $s_{12} \lesssim m_N^2$, we conclude that the off-shell production with the current assignments receives a further suppression roughly by a factor of $p_{\rm max}^2/m_{N}^2$. 
\subsection{CP-violation in the on-shell production}

The CP-violation for the on-shell production proceeds via very similar lines with the only difference that the loop functions now read 
\begin{align} 
f^{(H L)}_{ij }(x) &\equiv 2\int \frac{d^4p}{(2\pi)^4}\frac{1}{(p^2-  i\epsilon)((p-p_{\chi})^2 - i\epsilon)(p_{N}^2-m_{j}^2-i \epsilon)} \frac{\Tr \left[ \slashed{p}_N 
\slashed{p}_\chi \slashed{p}_N   \slashed{p} P_L\right]}{\Tr \left[ \slashed{p}_N \slashed{p}_\chi P_L  \right]}
\\
f^{(\chi \phi)}_{ij}(x) &\equiv \int \frac{d^4p}{(2\pi)^4}\frac{1}{(p^2-  i\epsilon)((p-p_{\chi})^2 - i\epsilon)(p_{N}^2-m_{j}^2-i \epsilon)} \frac{\Tr \left[ \slashed{p}_N 
\slashed{p}_\chi \slashed{p}_N   \slashed{p} P_L\right]}{\Tr \left[ \slashed{p}_N \slashed{p}_\chi P_L  \right]}
\end{align}
leading to the imaginary part of the form 
\begin{align}
\label{eq:Im_part_f_2}
& \text{Im}[f^{(HL)}_{ij}(x)] =\frac{1}{16\pi} \frac{1}{1-x_{ij}}, \qquad \text{Im}[f^{(\chi \phi)}_{ij}(x)] = \frac{1}{32\pi} \frac{1}{1-x_{ij}}  \, , \qquad  \qquad  x_{ij} = \frac{m_{j}^2}{m_{i}^2} \, . 
\end{align}
From this result, we observe that the asymmetry produced via this assignment is suppressed by a further factor of $m_i/m_j$ with respect to the scenario considered in the main text. Focusing on the regime where $m_1 \ll \gamma_w v \ll m_2 $ for the masses of $N_1$ and $N_2$ we then see that the CP-violation in production is suppressed by a factor of $\frac{m_1}{m_2}$ for the chirality assignment in Eq.\eqref{eq:main_model_1} as compared to that in Eq.\eqref{eq:main_model}. Similar conclusions can also be drawn about the CP-violation in the decay of the heavy fermion $N$, and thus we conclude that the contribution to CP-violation of the charge assignment in Eq.\eqref{eq:main_model_1} is subdominant to that of Eq.\eqref{eq:main_model}.

\subsection{Conclusion}

We conclude that in general, the chirality assignments discussed in this Appendix show a further suppression in the CP-violation and in the off-shell production with respect to the ones discussed in the main text. Consequently, we do not discuss them further. 

\section{U(1) gauged symmetry for the dark sector}
\label{app:other_real}

In this Appendix, we present an alternative realisation of the dark sector we discussed in the main text. We keep the same particle content, but now we introduce a new dark gauge symmetry $U(1)_D$ associated to a dark gauge boson $X^\mu$, and  assign the following $U(1)_D$ charges 
\begin{equation}
    q_{\phi} = -1,  \, \quad q_{\Tilde \chi} = 1, \, \quad q_{\chi} = 1 \quad  q_{\Tilde \phi} = 0, \, \quad q_{\rm SM} = 0 , \,   \quad q_N = 0 \, .
\end{equation}
In this setting, after $\phi$ gets a vev, the $U(1)_D$ symmetry is spontaneously broken, and the PT can then be associated with the breaking of this gauge symmetry. The Goldstone boson associated with the phase of $\phi$ will be absorbed by the dark gauge boson $X_{\mu}$ longitudinal mode, and so $X_{\mu}$ becomes massive. 
 In this setting, a growing pressure from the emission of soft gauge bosons of the form \cite{Bodeker:2017cim, Azatov:2020ufh, Gouttenoire:2021kjv, Azatov:2023xem, Ai:2025bjw}
 \bea 
 \mathcal{P}_{a\to b X^\mu } \propto \gamma_w \frac{g_X^3 v T_{\rm nuc}^3}{16\pi^2} \, 
 \eea 
 is unavoidable. Here $a,b$ are emitters coupling to $X$, that can be scalars, fermions or gauge bosons. This source of pressure leads to an upper bound on the terminal boost factor 
 \bea 
 \gamma^{\rm max}_w \sim \frac{16\pi^2 c_V}{g_X^3} \bigg(\frac{v}{T_{\rm nuc}}\bigg)^3  \, . 
 \eea 
 We thus impose $g_X \ll 1$ to still let the the bubble walls accelerate to very high velocities. In this framework, the list of allowed Yukawa interactions includes \footnote{For notational simplicity, we write this Lagrangian for only one family.}
 \bea 
 \mathcal{L}_{\rm yukawa} \supset y (\Bar{L}  H) P_R N + Y (\Bar{N} \phi) P_R \chi + y_{1} \Bar{\chi} (\Tilde \chi \Tilde \phi ) + y_3 \Bar{\chi} \chi \Tilde \phi + y_4 \Bar{\Tilde \chi} \Tilde \chi \Tilde \phi \, . 
 \eea 
Notice that importantly, the dangerous Yukawas of the form $y_{\Tilde \chi} \Bar{L} H P_R \Tilde \chi$ and $y_{\chi} \Bar{L} H P_R \chi$ are forbidden by the charge assignments. In this realisation,  the decay of the $\phi$ remnants is quick via $\phi \to XX$ and does not lead to any early matter domination. The asymmetry in the dark sector is again transmitted via the interaction $y_{1} \Bar{\chi} (\Tilde \chi \Tilde \phi )$, $\chi \to \tilde \chi \tilde \phi$. Freeze out of the $\tilde \chi$ symmetric abundance proceeds in the same way as in the former realisation: first from $\tilde \chi \to \tilde \phi$ via  $y_4 \Bar{\Tilde \chi} \Tilde \chi \Tilde \phi $ and then to the SM via the allowed interactions $\rho_1 \abs{H}^2 \Tilde \phi$ and $\lambda_{H \Tilde \phi} \abs{H}^2 \abs{ \Tilde \phi}^2$. The decay of the heavy dark photons can also be facilitated by a small kinetic mixing with the SM photon. 

The neutrino mass generation mechanism also proceeds in the same way with the explicit requirement that $\chi$ has to be a Dirac fermion. Importantly, the form of Eq.\eqref{eq:neutrino_mass} remains intact.

\section{Computation of the production from bubble collision: two methods, one result}
\label{app:production}

Production of particles due to the presence of a non-trivial classical field configuration requires modification of typical QFT Feynman rules.  In the literature different methods have been developed to estimate the abundance of heavy particles produced by the collision between the plasma and the non-trivial scalar condensate, that is the bubble wall; and between two bubble walls. More specifically, two different methods, respectively presented in~\cite{WATKINS1992446} and  in~\cite{Azatov:2021irb} (which we will from now on refer to as WW~\cite{WATKINS1992446} and AVY~\cite{Azatov:2021irb}) have been used in those two physically similar processes.  In this Appendix, we compare the two computations and show that they are actually equivalent. In the main text, we have used the AVY approach to compute loop correction and the related CP-violation.

As a toy model for the computation, we will consider the following interactions between the scalar condensate $\phi$ and a heavy Dirac fermion $\psi$
\bea 
\mathcal{L} = Y \phi \psi \bar \psi + \frac{1}{2}m_\psi \psi \bar \psi \, ,
\eea 
where $\psi$ is produced by the wall collision via $\phi \to \psi \bar \psi$. We will designate the condensate with $\varphi$ and the quantized scalar particles with $\phi$.  We will assume that $m_\psi \gg v$, so that the initial abundance of $\psi$ particles is vanishing and it is only produced by the bubble wall. 
 We first review the computation following the AVY approach and then show that the WW computation actually leads to the same result.

\subsection{AVY computation of the production}

 To compute the production of heavy particles, we start by computing the following  correlation function
$\langle 0| T\{ \bar \psi(x_1)\psi(x_2) \}|0\rangle$. We assume that the wall is located along the $x-y$ plane. The correlation function reads 
\bea
\langle 0| T\{ \bar \psi(x_1)\psi(x_2) \}|0\rangle= Y
\int d^4 x  \varphi(x) S_\psi(x_1-x| \varphi =0) S_\psi(x_2-x| \varphi =0) + \mathcal{O}\l(\frac{Yv}{m_\psi}\r)^2 \, ,
\eea
where we are expanding the correlation functions of the theory in the broken phase $v\neq 0$ in terms of the correlation functions $S_{\psi}(x, y| \varphi =0)$ of the unbroken $v=0$ phase. Such an expansion is called \emph{VEV Insertion Approximation} which was for example justified in \cite{Ai:2025bjw}. 
Defining the Fourier transform in the following way 
\bea 
\phi(x) = \int \frac{d^4p}{(2\pi)^4} e^{ipx} \phi(p)\, , 
\eea 
and going to momentum space, the correlation function becomes
\begin{align}
\langle 0| T\{ \bar \psi(x_1)\psi(x_2) \}|0\rangle & =Y\int \frac{d^4 x d^4 k d^4q}{(2 \pi)^8} e^{ik( x_1-x)+iq(x_2-x)
}S_\psi(k| \varphi =0)S_\psi(q | \varphi =0)\varphi(z, t)\nn
&=Y\int \frac{d^4 k d^4q}{(2 \pi)^8} e^{i k x_1 + i q x_2} S_\psi(k| \varphi =0)S_\psi(q | \varphi =0) 
\nn
&\times \l[(2\pi)^2 \delta^{(2)}(\vec{k}_\perp+\vec{q}_\perp) \int d zdt e^{i z(k_z+q_z) - i t (k_0+q_0)}
\varphi(z, t)\r]
 \, ,
\end{align}
where by definition 
\bea 
\delta^{(2)}(\vec{k}_\perp+ \vec{q}_\perp)\equiv \delta^{(1)}(k_x+ q_x)\delta^{(1)}(k_y+q_y)
\eea 
and in the second line, we assumed that all the energies involved in the transition are larger than the inverse length scale of the wall
\bea 
k_0 \approx q_0 \approx k_z \approx k_z' \gg 1/L_w,
\eea 
such that the Fourier transform can indeed be performed over a constant background.

Now we can use the LSZ reduction formula to relate the correlation function to the matrix element of the $\psi$ production $\langle \psi,k ; \psi^c, q \rangle$. We obtain
\bea
\langle \psi,k ; \psi^c, q \rangle= \underbrace{\l[(2\pi)^2 \delta^{(2)}(\vec{k}_\perp+\vec{q}_\perp)\int d zdt e^{i z(k_z+q_z) - i t (k_0+q_0)}
\varphi(z, t)\r]}_{\text{wall effect}}\times  \underbrace{\bar u_\psi(q)\ u_\psi(k) Y}_{\mathcal{M}_{\phi \to \psi \bar \psi}} \, .
\eea  
This expression has a nice interpretation as a factorisation of a \emph{wall effect} allowing the transition to occur and a traditional \emph{matrix element}, $\mathcal{M}_{\phi \to \psi \bar \psi} .$
Squaring this amplitude and summing over final spins, one obtains
\bea
|\langle \psi,k ; \psi^c, q \rangle|^2 = 4|Y|^2 q \cdot k \times \l[(2\pi)^2 \delta^{(2)}(k_\perp+q_\perp)\int d z dt e^{i z(q_z + k_z) - it (k_0+q_0)} \varphi(z,t)
\r]^2 \, . 
\eea 
From this, we can integrate over the phase space to obtain the probability of production 
\bea
P^{\rm AVY}_{\phi \to \psi \bar \psi}=\int \frac{d^3 qd^3 k}{(2\pi)^6 4q_0k_0}(2\pi)^2 \delta^{(2)}(k_\perp+q_\perp)|{\cal M}_{\phi \to \psi \bar \psi}|^2\l|\int dzdt e^{i z (k_z+ q_z)- it (k_0+q_0)}\varphi(z,t)\r|^2
\eea
or, after integrating over the perpendicular momenta \\
\bea
P^{\rm AVY}_{\phi \to \psi \bar \psi}=\int \frac{dk_z d^3q}{(2\pi)^4  4q_0k_0} |{\cal M}_{\phi(k_z+q_z, q_0+k_0) \to \psi \bar \psi}|^2\l|\int dzdt e^{i z (k_z+ q_z)- it (q_0+k_0)}\varphi(z,t)\r|^2 \, .
\eea
We first focus on the wall part, with the Fourier transform given by
\bea 
\tilde  \varphi(p) = (2\pi)^2\delta(p_x) \delta (p_y) \tilde \phi (p_z, \omega), \, 
\eea 
and the wall part of the production probability simplifying to  
\begin{align} 
\bigg|\int dzdt e^{i z (k_z+ q_z)- it (q_0+k_0)}\varphi(z, t)\bigg|^2 &= \bigg|\int \frac{d^4p}{(2\pi)^4} dzdt e^{i z (k_z+ q_z)- it (q_0+k_0)} e^{+ ipx}\tilde \varphi(p)\bigg|^2
\notag 
\\
&= \bigg|\int \frac{d^4p}{(2\pi)^4} (2\pi)^2 \delta(p_z-k_z- q_z)\delta(p_0 - q_0 - k_0) e^{-ip_\perp x_\perp} \tilde \varphi(p)\bigg|^2
\notag 
\\
&= \big|\tilde \phi(p_z =  k_z + q_z, \omega =  q_0 + k_0)\big|^2 \, . 
\end{align} 
Thus, we obtain 
\begin{align}
\label{eq:AVY_compu}
P^{\rm AVY}_{\phi \to \psi \bar \psi} &=\int \frac{dk_z d^3q}{(2\pi)^4  4q_0k_0} |{\cal M}_{\phi( k_z + q_z, q_0 + k_0) \to \psi \bar \psi}|^2 \big|\tilde \phi(p_z =  k_z + q_z, \omega = q_0 + k_0)\big|^2 
\notag\\ 
 &=\int \frac{dk_z d^2q_\perp dq_z}{(2\pi)^4  4q_0k_0} |{\cal M}_{\phi( k_z + q_z,  q_0 + k_0) \to \psi \bar \psi}|^2 \big|\tilde \phi(p_z =  k_z + q_z, \omega = q_0 + k_0)\big|^2 \, . 
\end{align}
The wall and the squared matrix element are independent of the $x-y$ directions. We can thus directly perform the $q_x, q_y$ integration in cylindrical coordinates: 
\bea 
\int^{}_0 \frac{d^2 q_\perp}{q_0 (\omega - q_0)} = \int^{}_0 \frac{\pi d q_\perp^2}{q_0 (\omega - q_0)} =  \int^{}_0 \frac{2\pi d \omega}{\omega} \, . 
\eea 
The probability of emission then becomes
\begin{align}
P^{\rm AVY}_{\phi \to \psi \bar \psi} & = Y^2\int \frac{\pi dk_z dq_z d\omega}{(2\pi)^4  4\omega }   (8k\cdot q - 8m_\psi^2)\big|\tilde \phi(p_z = k_z+ q_z, \omega = q_0+k_0)\big|^2 
\notag \\
&= Y^2\int \frac{\pi dp_z dP d\omega}{(2\pi)^4  4\omega }  (2\omega^2- 2 p_z^2 - 8 m_\psi^2) \big|\tilde \phi(p_z = k_z+ q_z, \omega = q_0+k_0)\big|^2  \,  ,
\end{align}
where we performed the change of variables $P = k_z-q_z, p_z = k_z+q_z$. We can therefore integrate over $P$ in the range \bea 
P/\omega  \in \big[- \sqrt{1  - 4m_\psi^2 /(\omega^2 -p_z^2)},\sqrt{1  - 4m_\psi^2/(\omega^2 -p_z^2)} \big] \, .
\eea 
The production probability can thus be simplified to 
\begin{align}
\label{eq:AVY_fermion_emission}
P^{\rm AVY}_{\phi \to \psi \bar \psi} 
=Y^2\int \frac{dp_zd\omega}{(2\pi)^2 4\pi  }  \frac{(\omega^2-p_z^2 - 4m_\psi^2)^{3/2}}{\sqrt{\omega^2- p_z^2}}\big|\tilde \phi(\omega^2 - p_z^2)\big|^2  \, . 
\end{align}

Finally, to arrive at the number of particles produced per unit surface, one needs to multiply by the number of emitted particles per interactions: 
\begin{align}
\label{eq:AVY_fermion_emission_particles}
\frac{N_\psi}{A}\bigg|^{\rm AVY}_{\phi \to \psi \bar \psi} 
&= 2Y^2\int \frac{dp_zd\omega}{(2\pi)^2 4\pi  }   \frac{(\omega^2-p_z^2 - 4m_\psi^2)^{3/2}}{\sqrt{\omega^2- p_z^2}} \theta (\omega^2-p_z^2 - 4m_\psi^2)\big|\phi(\omega^2 - p_z^2)\big|^2  \, . 
\end{align}
\subsection{WW computation}
On the other hand, the WW computation (which can be followed step by step in Ref.~\cite{WATKINS1992446}) yields a result of the form
\bea 
\label{eq:WW}
P^{\rm WW}_{\phi \to  \psi \bar \psi}= 2 \int \frac{dp_z d\omega}{(2\pi)^2} |\tilde \phi^2(p_z, \omega)|\text{Im}[\Sigma_{\phi \to  \psi \bar \psi}(\omega^2 - p_z^2)], 
\eea 
\bea 
\text{Im}[\Sigma_{\phi \to  \psi \bar \psi}(p^2)] = \frac{1}{2} \int\frac{d^3q d^3k}{(2\pi)^6 2E_k 2E_q }   \big|\mathcal{M}^2_{\phi \to \psi \bar \psi}\big| (2\pi)^4 \delta^{(4)} (p-k-q) \, .
\label{eq:prob_ww}
\eea 
When considering the decay of $\phi$ into two identical fermions, the following expression can be derived from Eq.\eqref{eq:prob_ww}

\begin{equation}
\text{Im}[\Sigma_{\phi \to  \psi \bar \psi}(p^2)]  = \frac{Y^2}{8\pi}\frac{(\omega^2-p_z^2 - 4m_\psi^2)^{3/2}}{\sqrt{\omega^{2}-p_z^2}} \, . 
\end{equation}
And satisfactorily, the number of particles per unit area computed in the WW method is given by 
\begin{align}
\label{eq:WW_fermion_emission}
\frac{N_\psi}{A}\bigg|^{\rm WW}_{\phi \to \psi \bar \psi} 
&= 2\int \frac{dp_zd\omega}{(2\pi)^2 4\pi  } Y^2  \frac{(\omega^2-p_z^2 - 4m^2)^{3/2}}{\sqrt{\omega^2- p_z^2}} \theta (\omega^2-p_z^2 - 4m^2)\big|\tilde \phi(\omega^2 - p_z^2)\big|^2  \, . 
\end{align}

\section{Computation of the Fourier transform of the wall}
\label{app:f_funct}

In this Appendix, we remind the basic formulae to compute the Fourier transform of the bubble wall collision, which we called $f$ function in the main text. Several papers attempted to estimate this function~\cite{Falkowski:2012fb,Mansour:2023fwj,Shakya:2023kjf}. Here, we present the results obtained from the numerical simulations in \cite{Mansour:2023fwj,Shakya:2023kjf}.

First of all, let us remind that two qualitatively different types of collisions have been analytically and numerically studied: the elastic and the inelastic collisions. In the former case, occurring when the minima are (almost) degenerate, the two
colliding bubble walls reflect off each other several times. The energy of the collision is used to re-establish the false vacuum in the region between
the receding walls. In the second instance, the energy of the collision is directly converted in scalar waves and the false vacuum is never re-established.

For a perfectly elastic collision, i.e. when the walls bounce back with the same relative speed and there is no
energy dissipation in scalar waves, the efficiency factor can be computed analytically~\cite{Falkowski:2012fb}:
\begin{equation}
    f_{\rm PE}(\chi) \equiv \frac{16 v^2}{\chi^2} \times \log\left[ \frac{2(\gamma_w/L_w)^2 - \chi +  \frac{2 \gamma_w}{L_w} \sqrt{\left( \frac{\gamma_w}{L_w} \right)^2 - \chi}}{\chi} \right] \Theta\bigg[(\gamma_w/L_w)^2 -\chi \bigg] \, , 
    \label{Falkowski:fPE}
\end{equation}
where $L_w$ denotes the length-scale of the wall in the wall frame, and $\chi \equiv p^2$ denotes the squared 4-momentum of the off-shell $\phi$. 

As for the case of elastic and inelastic collision the fit functions for the Fourier transform of the wall were obtained by the authors of \cite{Mansour:2023fwj} and are summarized below.
\begin{enumerate}
    \item {\textbf{Elastic collision numerically:}}
    In the case of a purely elastic BC, a good fit of the numerical results is provided by 
    \begin{equation}
        f_{\rm elastic}(\chi) = f_{\rm PE}(\chi) + \frac{v^2 L_{p}^2}{15 (m^{\rm true}_\phi)^2} \exp \left( - \frac{(\chi-(m^{\rm true}_\phi)^2 + 12 m^{\rm true}_\phi/L_p )^2}{440 (m^{\rm true}_\phi)^2/L_p^2} \right) \, , 
    \end{equation}
where we denote the mass of the scalar field $\phi$ around the true vacuum by $m_\phi^{\rm true}$. 
    The second term captures the contribution due to the frequency of oscillations around the false minimum, producing a peak in the Fourier transform. This peak becomes more and more Gaussian for larger values of space-time $L_p$ considered. In those expressions, one defined
    \begin{equation}
        L_p = \min \left( R_{\rm coll}, \Gamma^{-1} \right) \, , 
    \end{equation}
    where $R_{\rm coll}$ is the radius of the bubble at collision and $\Gamma^{-1}$ is the inverse decay rate of the scalar waves that gain energy from the oscillations. Notice that for $\chi \gg m_\phi^{\rm true}$ only the $f_{\rm PE}$ term contributes because the oscillation around the true minima is exponentially suppressed. This will be the relevant contribution in our study.

    \item {\textbf{Inelastic collision numerically:}}
    In the case of a totally inelastic collision, the numerical fit becomes
    \begin{equation}
        f_{\rm inelastic} = f_{\rm PE}(\chi) +\frac{v^2 L_{p}^2}{4 (m^{\rm false}_\phi)^2} \exp \left( - \frac{(\chi-(m^{\rm false}_\phi)^2 + 31 m^{\rm false}_\phi/L_p )^2}{650 (m^{\rm false}_\phi)^2/L_p^2} \right) \, . 
    \end{equation}
    Here, one again encounters the $f_{\rm PE}$ function already introduced for the elastic collisions, while the second term describes the oscillation around the false vacuum characterised by the mass $m_\phi^{\rm false}$.

\end{enumerate}

In the main text, we will always neglect the contribution from the peak, either elastic or inelastic, and keep the $f_{\rm PE}(\chi)$ piece. This is a conservative choice for the production mechanism motivated by the hierarchy $p \sim m_N \gg m_{\phi}^{\rm true},m_{\phi}^{\rm false}$. In other words, the oscillations around the peak can be neglected, and the Fourier transform in our case is simply given by $f_{\rm PE}(\chi)$ regardless of whether we use elastic or inelastic collisions. 

\begin{figure}[h!]
\centering
\subfigure{
\includegraphics[scale=0.265]{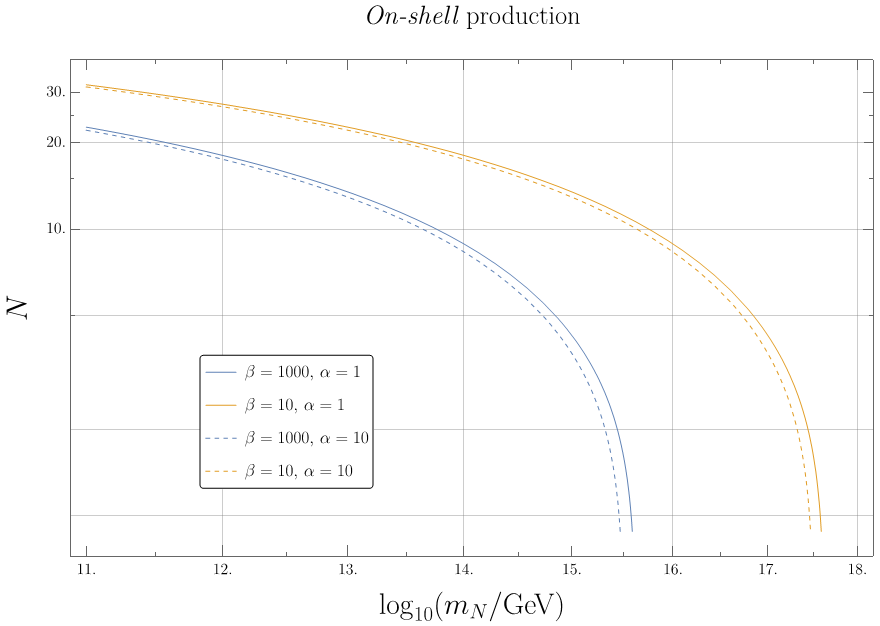}}
\hspace{2mm}
\subfigure{
\includegraphics[scale=0.265
]{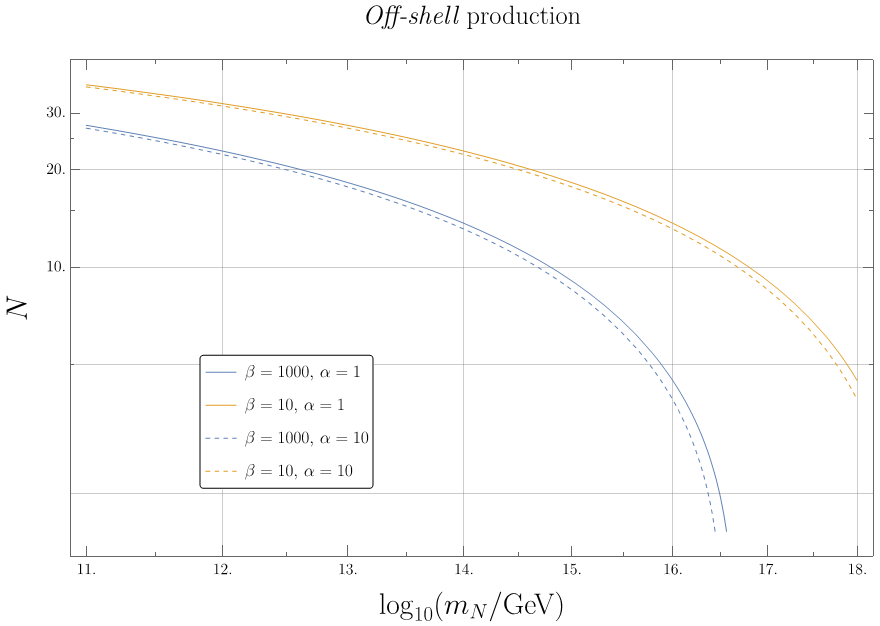}}
\caption{Evaluation of the function $N=2 \log{(\sqrt{z^2-1}+z)}$, with $z=  (2 \gamma_w v)/(m_{\chi} + m_N)$ for \textit{on-shell} $N$ production and $z= (2 \gamma_w v)/m_{\chi}$ for \textit{off-shell} one. We set $m_{\chi}= m_N/10$, $c_V=1$.} 
\label{fig:Nfactor}
\end{figure}

To simplify the numerics further, we assume that the integral of the efficiency factor for elastic collisions can be written as follows
\bea
    \int_{\chi_{\rm min}}^{\chi_{\rm max}} d\chi f_{\rm PE}(\chi) \simeq N(z=\sqrt{\chi_{\rm max}/\chi})\bigg|^{z_{\rm max}}_{z_{\rm min}} \times  \int_{\chi_{\rm min}}^{\chi_{\rm max}}  d\chi  \frac{16 v^2} {\chi^2} \, ,
\eea
where $N= 2 \log{(\sqrt{z^2-1}+z)}$. $N$ is evaluated at the extrema $z_{\rm max}= \sqrt{\chi_{\rm max}/ \chi_{\rm min}}$ and $z_{\rm min}=\sqrt{\chi_{\rm max}/ \chi_{\rm max}}=1$. Let us remind that $\chi_{\rm max}= (2 \gamma_w v)^2$ and in our scenario $\chi_{\rm min}= (m_{\chi}+m_N)^2$  for the \textit{on-shell} $N$ production and $\chi_{\rm min}= m_{\chi}^2$  for the \textit{off-shell} one. The function $N$ is plotted in Fig.\ref{fig:Nfactor} as a function of the $N$ mass $m_N$.

\bibliographystyle{JHEP}
{\footnotesize
\bibliography{biblio}}
\end{document}